\documentclass[]{elsarticle}

\journal{Journal of Computational Physics}

\usepackage{mathrsfs,amsfonts,amsmath,amssymb,epsfig,graphicx,algorithm,cancel, subfig}
\usepackage{booktabs, wrapfig}
\usepackage{epstopdf}
\usepackage{array}
\usepackage{xcolor}
\usepackage{hyperref}
\usepackage[margin=1.2in]{geometry}
\newcommand{\dds}[1]{\frac{d #1}{d s}}
\newcommand{\dd}[2]{\frac{d #1}{d #2}}
\newcommand{\pp}[2]{\frac{\partial #1}{\partial #2}}
\newcommand{\avg}[1]{\left\langle #1 \right\rangle}

\newcommand{\R}{\mathbb{R}}

\newcommand{\mus}{{m_{us}}}

\newcommand{\integrate}{\int_{0}^{T}}
\newcommand{\tend}{T}

\DeclareMathOperator*{\argmin}{arg\,min}

\newcommand{\nax}[1]{\textcolor{red}{#1}}
\newcommand\naxout{\bgroup\markoverwith{\textcolor{red}{\rule[0.5ex]{2pt}{0.4pt}}}\ULon}

\begin{document}

\begin{frontmatter}

\title{Sensitivity analysis on chaotic dynamical systems by Non-Intrusive Least Squares Shadowing (NILSS)}

\author[bkl]{Angxiu Ni}\corref{cor}
\ead{niangxiu@gmail.com}
\author[mit]{Qiqi Wang}
\ead{qiqi@mit.edu}

\cortext[cor]{Corresponding author.}
\address[bkl]{Mathematics, University of California, Berkeley, CA 94720, USA}
\address[mit]{Aeronautics and Astronautics, MIT, 77 Mass Ave, Cambridge, MA 02139, USA}

\begin{abstract}

This paper develops the non-intrusive formulation of the Least-squares shadowing (LSS) method,
for computing the sensitivity of long-time averaged objectives in chaotic dynamical systems.
This non-intrusive formulation constrains the computation to only the unstable subspace,
greatly reducing the cost of LSS for many problems;
moreover, it reparametrizes the LSS problem, requiring only minor modifications to existing tangent solvers. 
NILSS is demonstrated on a chaotic flow over a backward-facing step simulated with a mesh of $12\times10^3$ cells.
\end{abstract}


\end{frontmatter}

\noindent{}
\nax{(This is a revision of the published JCP version.
  Major changes include: 
  new abstract, new references, adding another derivation, 
  removing the previous incomplete discussion of FD-NILSS since it is given in detail in \cite{Ni_fdNILSS},
  new subscript numbering consistent with later papers.
  Edited \today.)}

\section{Introduction}

Many important phenomena in engineering, such as turbulent flow \cite{Kolmogorov_turbulence}
and some fluid-structure interactions \cite{Dowell1982}, are chaotic. 
In these systems, the objectives we are often interested in are long-time averaged rather than instantaneous quantities.
Furthermore, we want to perform sensitivity analysis, that is,
we want to know how a change in the parameters of a system can affect its objectives.
Such sensitivity analysis is the purpose of this paper.

To rigorously define the problem, we first consider the governing equation for a chaotic dynamical system: 
\begin{equation} \label{eq:dynamical system}
\dd{u}{t} = f(u,s), \quad u(t=0) = u_0(\phi),
\end{equation}
where $f(u,s):\R^m\times \R\rightarrow\R^m$ is a smooth function, $u$ is the state, and $s$ is the parameter. 
The initial condition $u_0$ is a smooth function of $\phi$.
A solution $u(t)$ is called the primal solution.

In this paper, The objective is a long-time averaged quantity.
To define it, we first let $J(u,s):\R^m\times\R\rightarrow\R$ be a continuous function that represents the instantaneous objective function.
The objective is obtained by averaging $J$ over a infinitely long trajectory:
\begin{equation} \label{eq:average J}
\avg{J}_\infty:= \lim\limits_{t\rightarrow\infty}\avg{J}_T, \text{ where }\avg{J}_T:= \frac{1}{T}\integrate J(u,s)dt.
\end{equation}
$ \avg{J}_T $ depends on $s$, $\phi$, and $T$, while $ \avg{J}_\infty $ is determined only by $s$ and $u_0$.
Here we make the assumption of ergodicity \cite{walters2000introduction}, which means that $u_0$, hence $\phi$ does not affect $\avg{J}_\infty$.
As a result, $ \avg{J}_\infty $ only depends on $s$.

The purpose of this paper is to develop an algorithm that computes the sensitivity $d\avg{J}_\infty / ds$.
The sensitivity can help scientists and engineers design products \cite{Jameson1988,Reuther}, 
control processes and systems \cite{Bewley2001,Bewley2001a}, solve inverse problems \cite{Tromp}, 
estimate simulation errors \cite{Becker2001,Giles2002,Fidkowski}, 
assimilate measurement data \cite{Thepaut1991,COURTIER1993}, quantify uncertainties \cite{Wang_ODE_LSS},
and train neural networks \cite{deeplearning_book_Goodfellow,linearRange_GD}.

For chaotic dynamical systems, computing $d \avg{J}_\infty / ds $ is challenging, since in general:
\begin{equation}\label{eq:commutelimit}
	\dds{}\avg{J}_\infty \ne \lim\limits_{T\rightarrow\infty}\pp{}{s}\avg{J}_T (s, \phi, T).
\end{equation}
That is, if we fix $u_0(\phi)$, the process of $T\rightarrow \infty$ does not commute with differentiation with respect to $s$.
As a result, the transient method, which employs the conventional tangent method with a fixed $u_0$,
does not converge to the correct sensitivity for chaotic systems. 
In fact, the transient method diverges most of the time \cite{Blonigan_lss_airfoil}.

Many sensitivity analysis methods have been developed to compute $d\avg{J}_\infty / ds$.
The conventional methods include the finite difference and transient method.
The ensemble method, developed by Lea et al. \cite{Lea2000,eyink2004ruelle}, 
computes the sensitivity by averaging results from the transient method over an ensemble of trajectories.
Another recent approach is based on the fluctuation dissipation theorem (FDT), as seen in \cite{Thuburn2005, Palmer2001, young2002srb,Leith1975,abramov2007blended, Abramov2008}.

In this research study, we consider the least squares shadowing (LSS) approach, 
developed by Wang, Hu and Blonigan \cite{wang2014convergence,Wang_ODE_LSS}. 
LSS computes a bounded shift of a trajectory under an infinitesimal parameter change, which is called the LSS solution.
The LSS solution can then be used to compute the derivative $d \avg{J}_\infty / ds$.
LSS has been successfully applied to dynamical systems such as the Lorenz 63 system 
and a modified Kuramoto-Sivashinsky equation \cite{Blonigan2014,Wang_ODE_LSS,BloniganPhdThesis}. 
LSS has also been applied, by Blonigan et al., to sensitivity analysis for flow around airfoils  \cite{Blonigan_lss_airfoil}.
From a theoretical standpoint, Wang has proven that, under ergodicity and hyperbolicity assumptions,
LSS converges to the correct sensitivity at a rate of $ T^{-0.5} $, where $T$ is the trajectory time length\cite{wang2014convergence}.

However, for large systems which arise in real life problems, LSS is expensive, since it involves solving a large linear system, 
where the number of variables is the system dimension times the number of time steps.
As the system gets larger and the trajectory longer, the linear system becomes very large and possibly stiff.
Although solving the system could be accelerated by preconditioners and iterative methods \cite{BloniganPhdThesis},
there would still be a large cost in both computational time and memory.
Furthermore, LSS requires the Jacobian matrix $\partial _u f(u,s)$ at each time step,
which many existing simulation software may not readily provide; and making modifications to existing codes can be difficult.

To reduce the computational cost and ease the implementation of LSS, 
this paper develops the non-intrusive least squares shadowing (NILSS) method.
The computational and memory cost of NILSS are both proportional to the number of positive Lyapunov exponents (LE). 
For many real life applications this number is much smaller than the dimension of the dynamical system, 
and the cost of NILSS is much lower than LSS.
Another benefit is that NILSS requires less modification to the underlying tangent solver than LSS, 
since it does not require the Jacobian matrix $\partial_u f(u,s)$.

At the time of this revision, there are several updates for non-intrusive shadowing methods.
The covariant Lyapunov vectors (CLV) and shadowing directions of a three dimensional flow over a cylinder
was investigated in \cite{Ni_CLV_cylinder},
leading to the conjecture that large portions of CLVs in open flows are stable,
and hence non-intrusive formulation is important for achieving high efficiency.
The finite difference NILSS (FD-NILSS) \cite{Ni_fdNILSS} uses finite difference to approximate the tangent solutions used in NILSS,
hence it no longer requires tangent solvers.
Ni found an adjoint version of the shadowing lemma \cite{Ni_adjoint_shadowing},
based on which Ni and Talnikar developed NILSAS, the adjoint counterpart of NILSS \cite{Ni_nilsas}.

The rest of this paper is arranged as follows.
We start by giving two derivations of the non-intrusive formulation, in particular, 
the second derivation in section~\ref{s:short derivation from LSS} is newly added in this revision.
Next, we address some numerical issues in the NILSS algorithm.
Then, we present a step-by-step description of the NILSS algorithm.
Finally, we apply NILSS to the Lorenz 63 system and a CFD simulation of a flow over a backward-facing step.

\section{A derivation of the non-intrusive formulation}
\label{s:longer derivation of nilss}

We first give a derivation based on distilling the long-time effect by subtracting the transient effect.
In section~\ref{ss:2 perturbations}, we examine the long-time and transient effects due to perturbations in the system parameters;
and how transient effects are also generated by perturbations in initial conditions.
In section \ref{ss:two perturbations}, we mathematically define the two perturbations as homogeneous and inhomogeneous tangent solutions.
To distill the long-time effect, denoted by some inhomogeneous tangent $v$,
we want to construct a homogeneous tangent $w$ which represents the transient effect brought about by varying initial conditions,
and subtract it from the conventional inhomogeneous tangent $v^*$, which represents the two effects of varying parameters.
In section \ref{ss:w by CLV}, we see how to mathematically construct such a $w$
as a linear combination of unstable Characteristic Lyapunov Vectors (CLV).
In section \ref{ss:w by wj}, we give a computationally efficient formula for $w$ 
from only the conventional tangent $v^*$ and several homogeneous solutions $\{w_j\}$, which approximate unstable CLVs.
Finally, section \ref{ss:compute djds1} explains how to compute $d \avg{J}_\infty / ds$.

\subsection{Connection between sensitivity to system parameters and initial conditions} 
\label{ss:2 perturbations}

As we have seen in our definition of the objective in equation (\ref{eq:average J}), the average is taken over an infinitely long trajectory. 
The sensitivity of the objective could be revealed by looking at perturbations in the trajectory due to perturbations in the parameters.
Such perturbations are examined in this section.

Trajectories of chaotic dynamical systems depend sensitively on system parameters.
If we change any parameter by a small amount, the new trajectory will be significantly different than the old one, 
even though they start from the same initial condition.
This is similar to another sensitive dependence on initial conditions, better known as the `butterfly effect',
that is, a small difference in the initial condition can grow larger and larger as the system evolves.

\begin{figure}[!htb]
  \centering
  \begin{subfloat}
    {\includegraphics[trim=4cm 1cm 4cm 3cm, clip=true, width=0.47\textwidth]{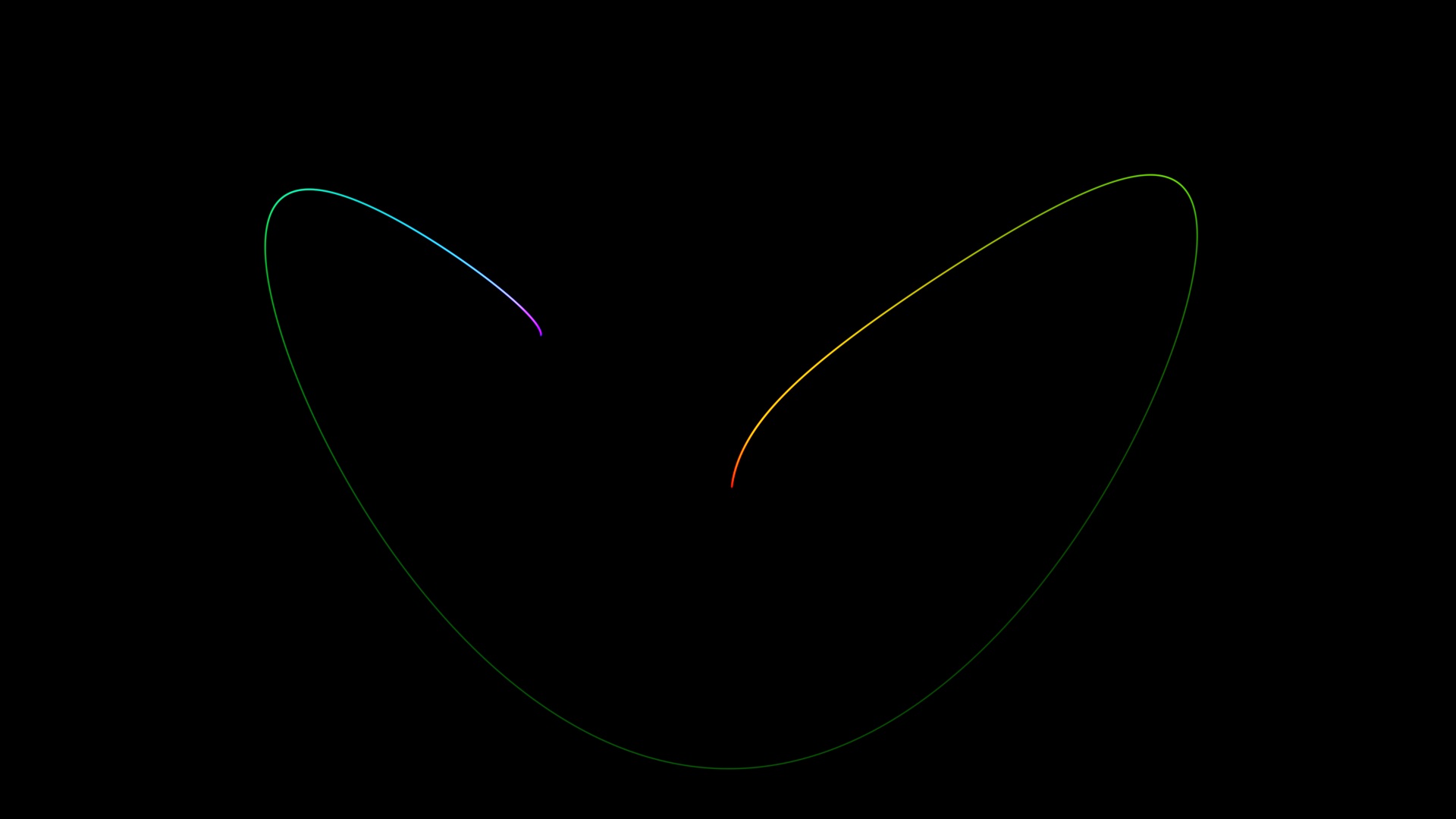}}
  \end{subfloat}
  \begin{subfloat}
    {\includegraphics[trim=4cm 1cm 4cm 3cm, clip=true, width=0.47\textwidth]{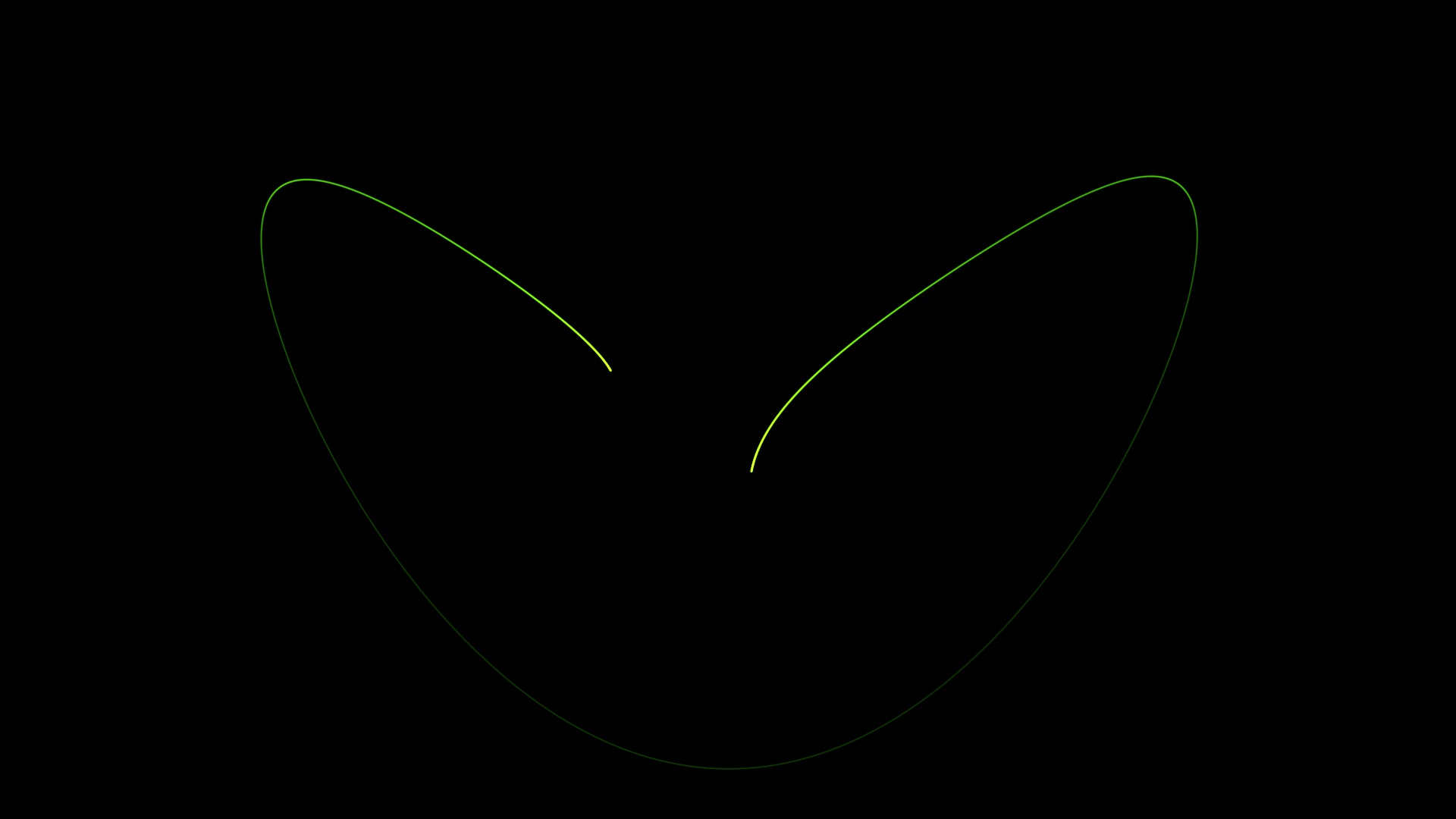}}
  \end{subfloat}\\
  \begin{subfloat}
    {\includegraphics[trim=4cm 1cm 4cm 3cm, clip=true, width=0.47\textwidth]{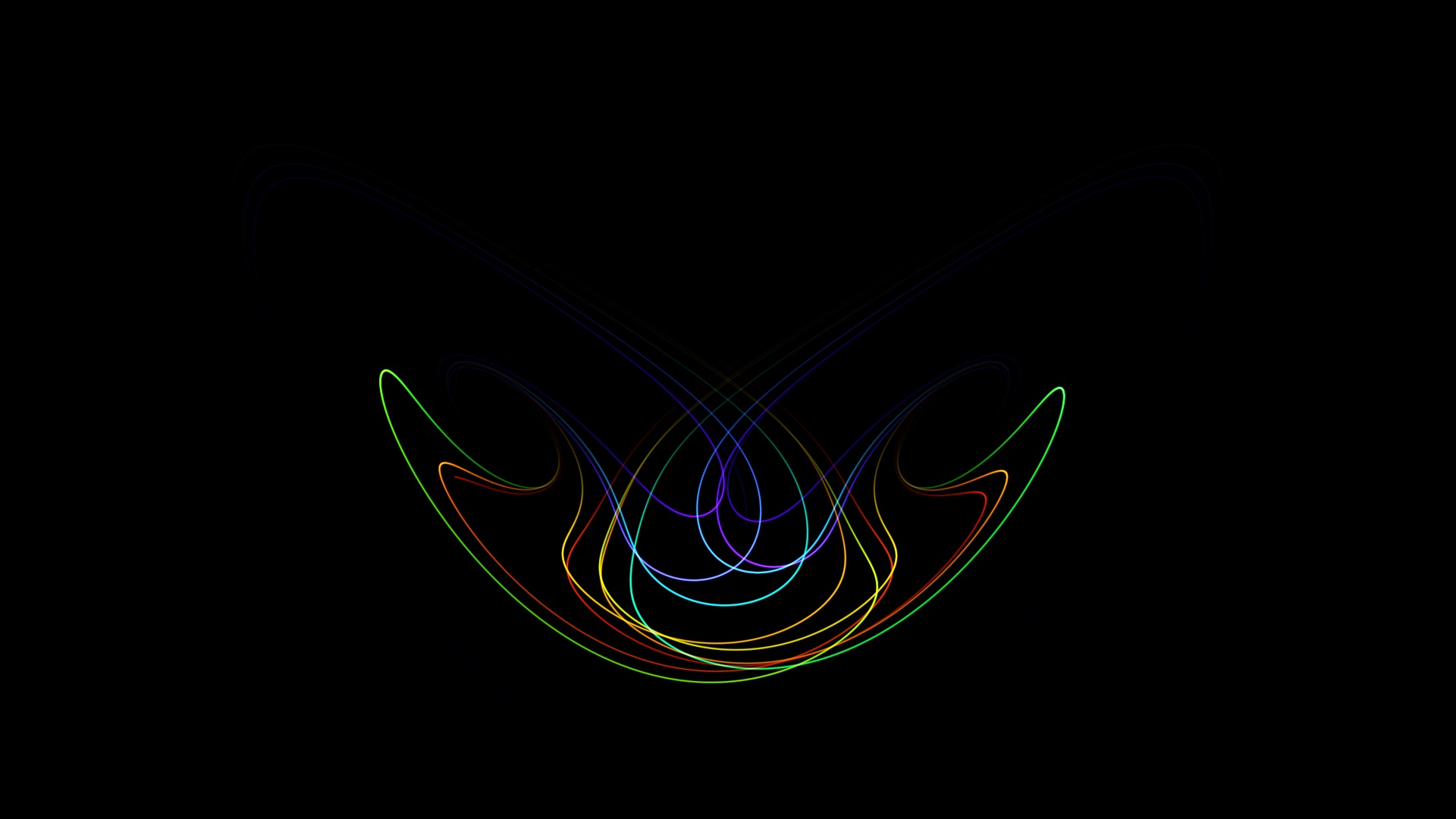}}
  \end{subfloat}
  \begin{subfloat}
    {\includegraphics[trim=4cm 1cm 4cm 3cm, clip=true, width=0.47\textwidth]{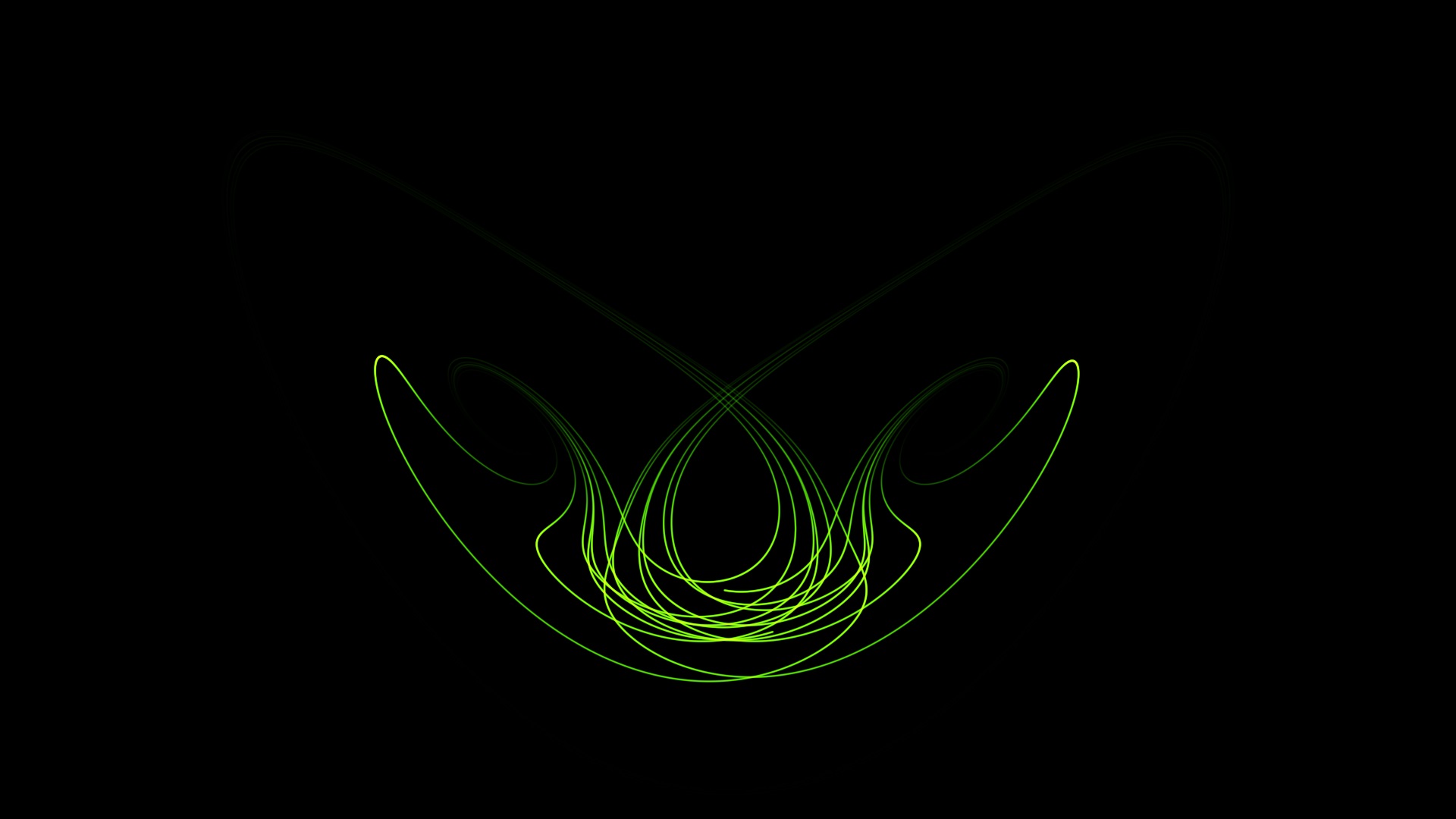}}
  \end{subfloat}\\
  \begin{subfloat}
    {\includegraphics[trim=4cm 1cm 4cm 3cm, clip=true, width=0.47\textwidth]{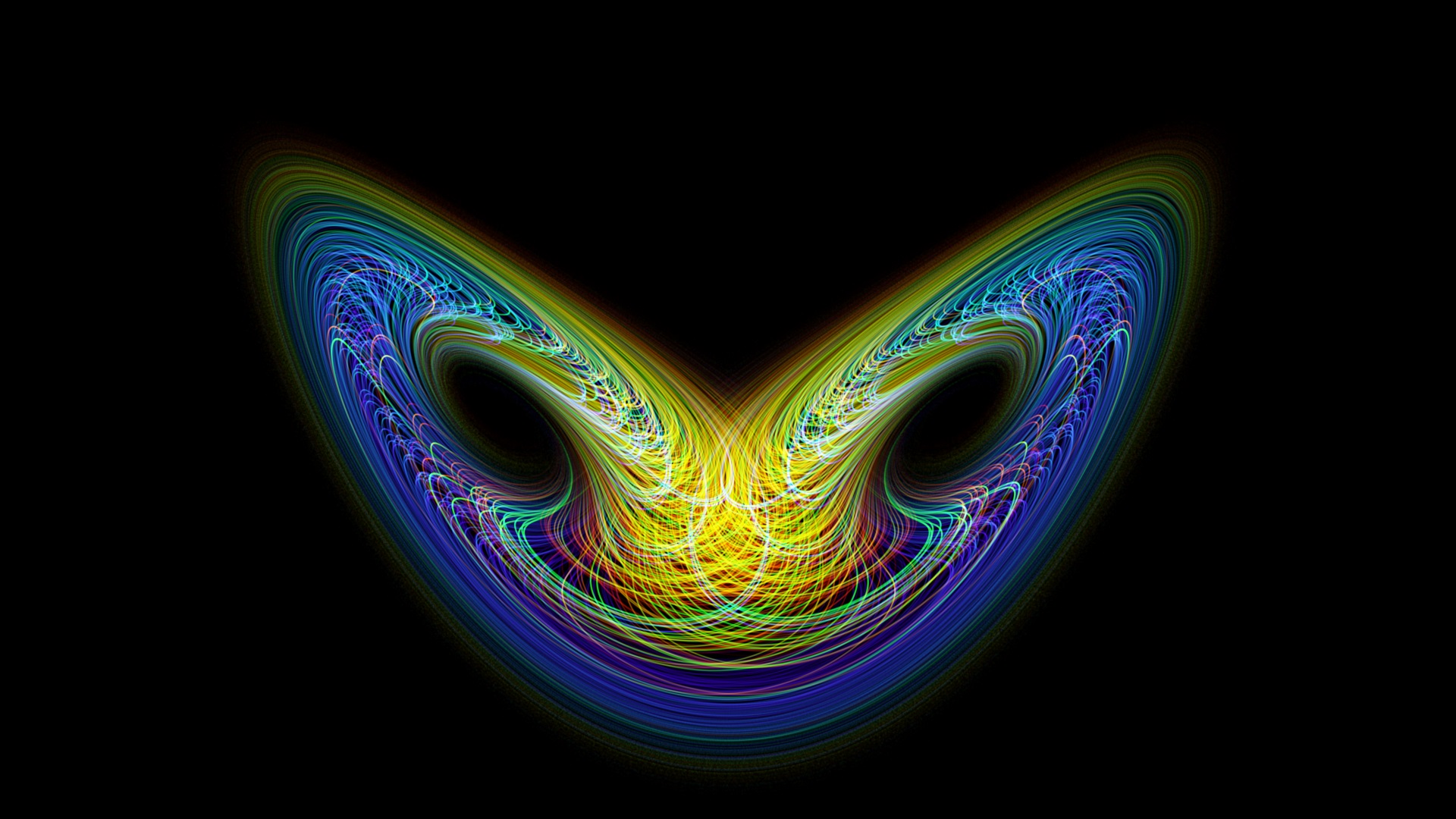}}
  \end{subfloat}
  \begin{subfloat}
    {\includegraphics[trim=4cm 1cm 4cm 3cm, clip=true, width=0.47\textwidth]{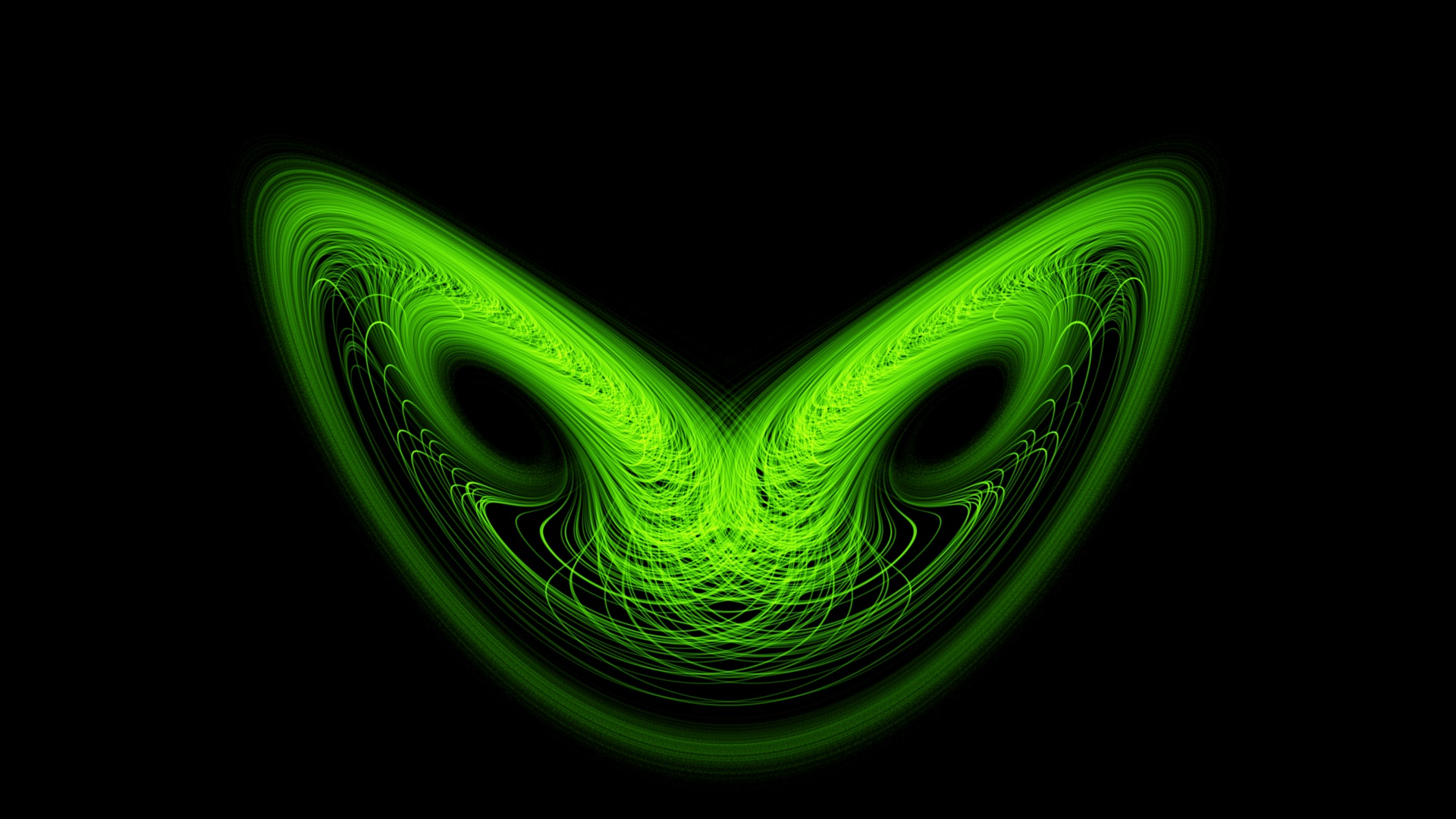}}
  \end{subfloat}\\
  \begin{subfloat}
    {\includegraphics[trim=4cm 1cm 4cm 3cm, clip=true, width=0.47\textwidth]{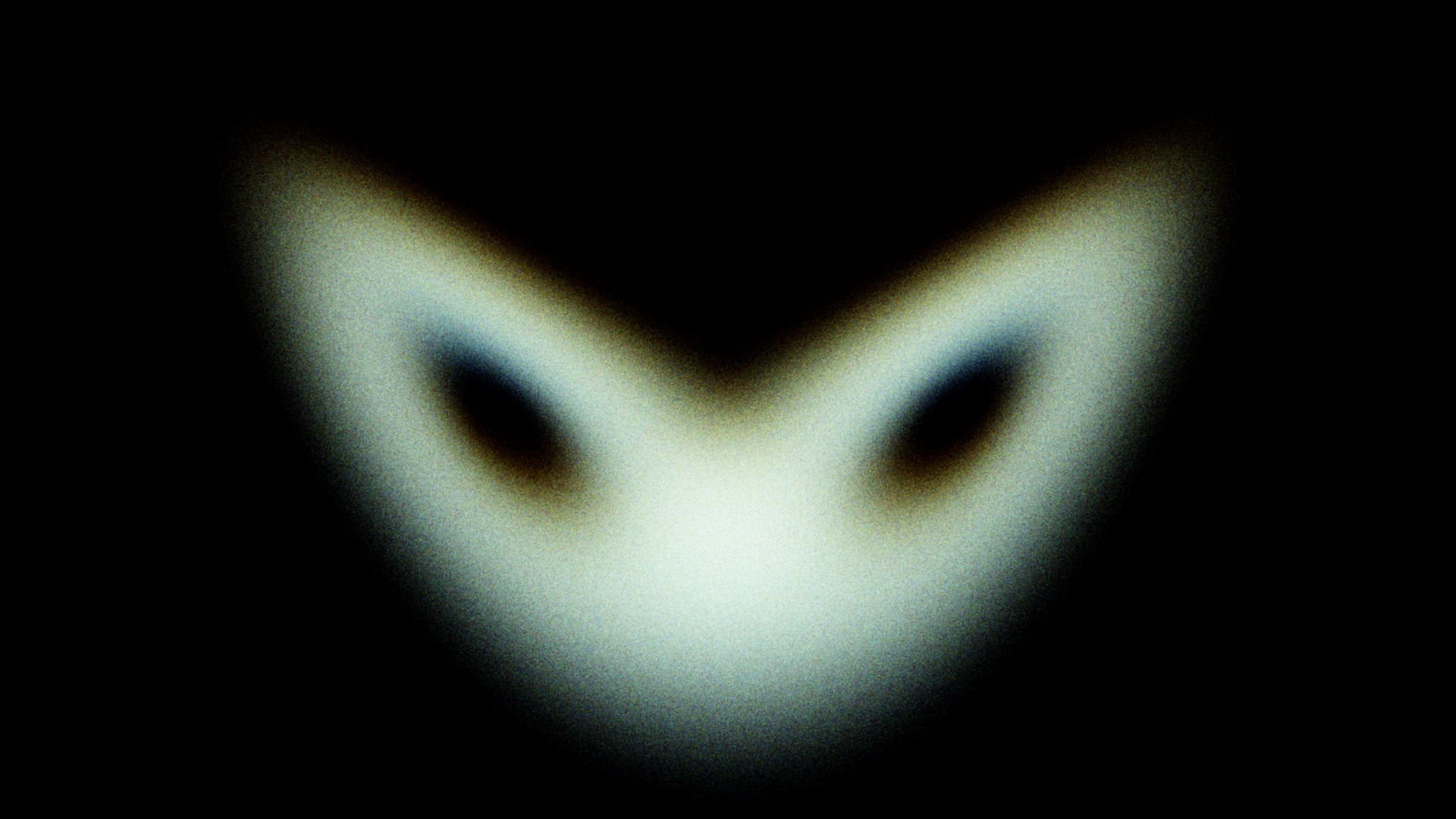}}
  \end{subfloat}
  \begin{subfloat}
    {\includegraphics[trim=4cm 1cm 4cm 3cm, clip=true, width=0.47\textwidth]{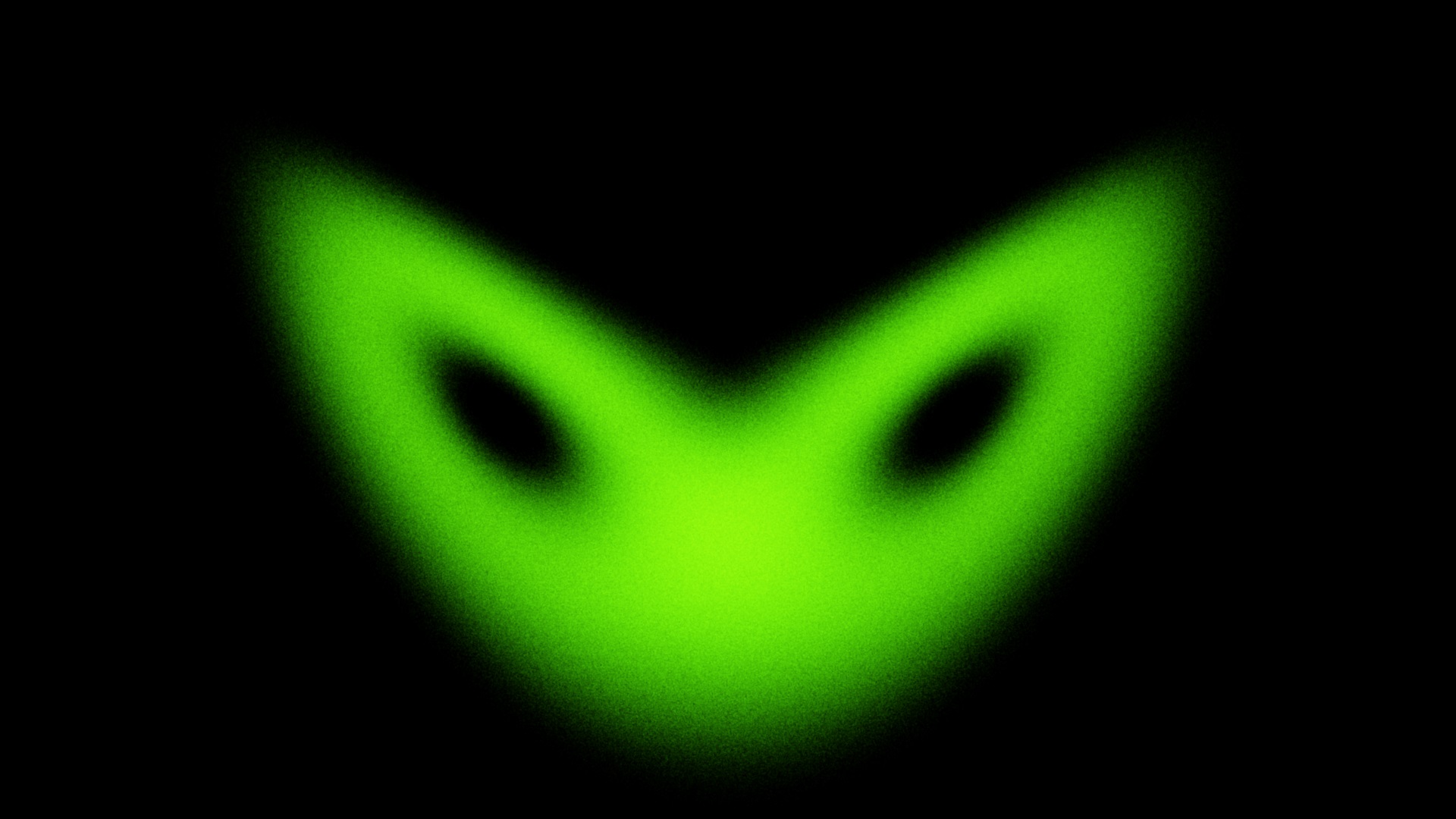}}
  \end{subfloat}\\
  \caption{Snapshots of an ensemble of $ 1.8\times 10^7$ trajectories of the Lorenz 63 system. 
  Left column: trajectories with different parameters that are uniformly distributed over the range [27,29], 
  where smaller $\rho$ is indicated by blue, larger $\rho$ by red.  
  Right column: trajectories with fixed $\rho$ but with initial conditions uniformly distributed over
  $(12.00, 6.82, 36.47)\pm[0.0939, -0.001053, 1.025]$. 
  From top to bottom: snapshots taken at time 1.67, 5.0, 10.0, and 41.67.}
  \label{fig:sensitiveDependence}
\end{figure}

To illustrate the similarity between the two sensitivities, we consider the Lorenz 63 system,
which is a simplified ODE model for atmospheric convection \cite{Lorenz1963}.
Lorenz 63 has three states $x, y, z$ and a parameter $\rho$.
In figure \ref{fig:sensitiveDependence}, we show the sensitive dependence of trajectories on both the initial condition and the parameter.
In the left column, on the $x$-$z$ axis,
we plot planar snapshots of $ 1.8\times 10^7 $ trajectories with varying $\rho$ but with the same initial condition $u_0 = (12.00, 6.82, 36.47)$.
Here $\rho$ is uniformly distributed in $28\pm \Delta \rho$, where $\Delta \rho=1$.
Note that a smaller $\rho$ is indicated by colors with shorter wavelengths (blue), while a larger $\rho$ by longer wavelengths (red).
On the right column, we plot snapshots of the same number of trajectories with the same parameter $\rho = 28$,
but with a varying initial condition, which is characterized by a vector that is uniformly distributed along $u_0\pm \Delta u_0$,
where $\Delta u_0 = [0.0939, -0.001053, 1.025]$.
As we shall see later, $\Delta u_0$ is chosen to have similar effects to the transient effect of varying $\rho$.

There are many similarities and subtle differences between the effects of a varying $\rho$ and a varying $u_0$.
As we can see in the first three rows of figure \ref{fig:sensitiveDependence}, in the short time, 
the varying $\rho$ results in diverging trajectories which looks like the effect of only varying $u_0$: 
we call this the transient effect.

The picture in the last row of figure \ref{fig:sensitiveDependence} is obtained after letting the trajectories evolve over a long time.
The picture on the right gives the attractor of the base parameter.
The picture on the left, at first glance, has similar shape as the attractor to its right. 
However, the left figure is the superposition of many attractors with different parameters,
and a closer look shows that it has different colors in different parts.
The red color on the upper rim, and the blue on the lower, indicates that as $\rho$ increases, the attractor moves upward in the $z$ direction.
To conclude, in the long-time, varying $\rho$ results in a shifted attractor: we call this the long-time effect.

The long-time effect generated by a varying $\rho$ is important for computing the long-time sensitivity,
however, it is hidden beneath diverging trajectories and is only visible after a long time and an ensemble of millions of trajectories.
As we said, the transient effect is reflected by diverging trajectories, hence if we can find two trajectories,
one with $\rho$ and another with $\rho+\delta \rho$, which do not diverge, then their difference does not contain the transient effect.
Now with the transient effect gone, their difference contains only the long-time effect.
Thus, we can reveal the long-time effect with a shorter trajectory.

Our main goal in this paper is to devise an algorithm that can generate the transient effect
and subsequently `subtract' the transient effect from a varying $\rho$, 
so that we can find two trajectories that do not diverge from each other, and whose difference only contains the long-time effect.
In fact, in figure \ref{fig:sensitiveDependence}, $\Delta u_0 = v\Delta \rho$ and $v$ represents the NILSS solution.
As we shall see, this change in the initial condition yields the transient effect,
and by subtracting it from the two effects of a varying $\rho$, we can distill the long-time effect using a short trajectory.
We will clarify the qualitative description of `subtraction' in later sections.

\subsection{Describing perturbations by tangents}\label{ss:two perturbations}

Now we mathematically describe the trajectory perturbations generated by parameter and initial condition perturbations. 
This is done by tangent solutions.
Specifically, the perturbation due to parameter change is described by inhomogeneous tangents, 
while that due to initial condition change is described by homogeneous tangents.

First, we differentiate the dynamical system in equation (\ref{eq:dynamical system}) with respect to $s$, while keeping $\phi$ fixed.
Then, we let $v^*= \partial u/\partial s$.
Thus, the governing equation for $v^*$ is: 
\begin{equation} \label{eq:v*}
  \dd{v^*}{t} - \partial_u f v^* = \partial_s f, \quad v^*(t=0)=0
\end{equation}
where $\partial_u f$ is an $\R^m\times \R^m$ matrix and $\partial_s f$ is an $\R^m$ column vector.
The zero initial condition $v^*$ reflects that $u_0$ remains unchanged.
By definition, $v^*$ reflects the trajectory perturbation due to parameter changes under fixed initial conditions, 
which is shown in the left column of figure \ref{fig:sensitiveDependence}.

Under the assumption of ergodicity, the long time behavior is not affected by the selection of initial conditions.
This suggests that $\phi$ is not necessarily fixed if we are only interested in the change of the long-time average.
We define inhomogeneous tangent solutions as solutions that satisfy the ODE in equation (\ref{eq:v*}), but without the initial condition:
\begin{equation} \label{eq:tangent equation}
  \dd{v}{t} - \partial_u f v = \partial_s f.
\end{equation}
Notice that this ODE is under-determined.
To get a solution we can either provide an initial condition, as we did for $v^*$, or put the ODE as a constraint for some optimization problems.
By its definition, $v$ reflects the trajectory perturbation due to a change in the parameter, while the initial condition change is not specified.

We define $w= \partial u / \partial \phi $, where $s$ is assumed to be fixed and $w$ satisfies the so called homogeneous tangent equation:
\begin{equation} \label{eq:homogeneous tangent equation}
  \dd{w}{t} - \partial_u f w = 0.
\end{equation}
By its definition, $w$ characterizes the perturbations due to initial condition changes while $s$ is fixed, 
as shown in the right column of figure \ref{fig:sensitiveDependence}.

Hence $v^*$ and $w$ describe the effect of only varying $s$ and $u_0$, respectively. 
Also, equation (\ref{eq:homogeneous tangent equation}) differs from equation (\ref{eq:tangent equation}) by setting the right hand side to zero.
For two different inhomogeneous tangent solutions, say $v^*$ and an arbitrary $v$, their difference is a homogeneous tangent solution $w$.

We know that if we vary $s$, we generate two effects: one is equivalent to varying $u_0$; while the other shifts the attractor. 
Since we are interested in the latter, we want to find a $w$ such that $v = v^*+w$ contains only the long-time but not the transient effect.
Here we used addition, but we can replace $w$ by $-w$ so that we have subtraction in the formula.

Subtracting such $w$ from $v^*$ is the main idea behind NILSS.
As discussed in section \ref{ss:2 perturbations}, we want to find two trajectories,
one associated with parameter $s$ and the other with $s+\delta s$, which do not diverge.
Given the tangent solution definition, we can mathematically state that a $v$, 
if its Euclidean norm\footnote{In this paper, the norm we use is Euclidean norm.} 
of its orthogonal projection onto $V^\perp(u)$ remains bounded as the trajectory length goes to infinity, 
then this $v$ suffices to reveal the long-time effect of the varying parameter.
We denote this sufficient $v$ by the shadowing direction, $v^\infty$, whose existence is proved by the shadowing lemma \cite{Pilyugin1999book}.
Here $V^\perp(u)$ is defined as:
\begin{equation}
  V^\perp(u)=\{p\in \R ^m: p^T f(u) = 0\},
\end{equation}
where $p^T$ is the transpose of the column vector $p$. Moreover, the orthogonal projection $p^\perp$ of $p$ is defined as:
\begin{equation}\label{eq:vperp projection}
  p^\perp = p - \frac{f^T p}{f^T f} f.
\end{equation}
$v^{\infty\perp}$ is defined by substituting $p$ by $v^\infty$. 
We define $w^\perp$, $\delta u^\perp$, $v^{*\perp}$, and $\{\zeta_j^\perp\}$ in a similar way.
We use the norm of $v^{\infty\perp}$ because it describes the perpendicular distance between two trajectories.
A more mathematical explanation of why such $v^\infty$ can be used to compute the sensitivity is in \ref{ss:derive djds}.

\subsection{Constructing \texorpdfstring{$w$}{w} from unstable Characteristic Lyapunov Vectors (CLV)} 
\label{ss:w by CLV}

The main goal of NILSS is to find a $w$ such that $v^\perp = v^{*\perp}+w^\perp$ approximates $v^{\infty\perp}$ on a finite trajectory.
Here $v^{\infty}$ is the inhomogeneous tangent the norm of whose projection, $v^{\infty\perp}$, remains bounded even on an infinitely long trajectory.
Notice that the NILSS solution $v^\perp$ may be not bounded if we extend it to an infinitely long trajectory; however, on the finite trajectory where NILSS is solved, $v^\perp$ provides a good approximation of $v^{\infty\perp}$.
Specifically, this means that if we apply both $v^\perp$ and $v^{\infty\perp}$ to the formula that computes sensitivity in equation (\ref{eq:dJds seg}), the results are similar.

In this subsection, we shall see one way to construct such a $w$ by supposing that we know $v^{\infty}$ and all CLVs.
This method is unrealistic since it requires too much computation.
Yet it is informative since it shows that we only need a linear combination of unstable CLVs to construct a desired $w$.
Based on this knowledge, we develop the NILSS method in the next subsection.

To further clarify this method, we should first define Lyapunov Exponent (LE) and the corresponding CLVs.
We assume that the dynamical system has a full set of LEs and corresponding CLVs \cite{Ruelle1979}.
That is, there are $\{\lambda_j, j=1,2,\cdots,m\}$, such that for any trajectory on the attractor and a corresponding homogeneous tangent solution $w(u)$, there is a unique representation of $w(u)$:
\begin{equation}\label{eq:subspace decomposition}
	w(u) = \sum_{j=1}^{m} a_j \zeta_j(u),
\end{equation}
where $a_j\in \R$ is a constant for all $u(t)$ on the trajectory.

Here each $\zeta_j(u)$ is a homogeneous tangent solution, and its norm behaves like an exponential function of time. 
That is, there exists $C_1,C_2>0$, such that for any $u(t)$ on the attractor and any $j$ and $t$, a CLV satisfies
\begin{equation}\label{eq:CLV}
	C_1 e^{\lambda_j t }\|\zeta_j(u(0))\|  \le \|\zeta_j(u(t))\| \le C_2 e^{\lambda_j t }\|\zeta_j(u(0))\| ,
\end{equation}
where $\{\lambda_j\}$ and $\{\zeta_j\}$ are LEs and CLVs, respectively.
CLVs with $\lambda_j>0$ are called unstable modes, those with negative $\lambda_j<0$ are stable modes, and those with $\lambda_j=0$ are neutral modes.
We denote the unstable modes by $\zeta_1, \cdots, \zeta_\mus$ and the neutral modes by $\zeta_m$. The remaining modes are the stable modes.
In fact, unstable modes are the reason for the `butterfly effect' since a perturbation in the unstable subspace grows exponentially over time.

We assume that for all $s$ we are interested in, there is no point $u$ on the attractor $\Lambda$ such that $f(u)=0$ and $\Lambda$ is bounded.
These two assumptions imply that, per \ref{ss:f is 0 CLV},  $f(u)$ is a CLV whose LE is 0.
We further assume that $f(u)$ is the only neutral mode.

Although CLVs are not necessarily in $V^\perp$, we can project them onto $V^\perp$.
Thus, equation (\ref{eq:subspace decomposition}) becomes:  
\begin{equation}\label{eq:Vperp decomposition}
w^\perp = \sum_{j=1}^{m-1} a_j \, \zeta^\perp_j(u) ,
\end{equation}
where $w^\perp$ and $\zeta^\perp_j$ are orthogonal projections as defined by equation (\ref{eq:vperp projection}).
Because $V^\perp$ is perpendicular to $f(u)$, the projection of the neutral mode is zero.
This implies that the summation in equation (\ref{eq:Vperp decomposition}) only considers the stable and unstable modes, the total number of which is $m-1$.
We also call $\zeta^\perp_j$ stable or unstable modes based on their corresponding $\lambda_j$.

We assume that all CLVs are uniformly bounded away from each other.
Under this assumption, the norm of stable and unstable modes $\{\zeta_j^\perp\}$ also behave like exponentials, as defined in equation \ref{eq:CLV}.
\ref{ss: CLV perp are exponentials} justifies this claim.

Suppose that $v^\infty$ and its CLVs are known.
Since  $v^{\infty}- v^{*}$ is a homogeneous tangent solution, we can decompose $v^{\infty\perp}- v^{*\perp}$ via equation (\ref{eq:Vperp decomposition}).
By using the first $\mus$ coefficients in this decomposition, we let
\begin{equation}
w =  \sum_{j =1}^{\mus} a_j \zeta_j .
\end{equation} 
Thus, $v^\perp = v^{*\perp} + w$ approximates $v^{\infty\perp}$ since $v^{\infty\perp} - v^\perp$ is composed of only stable modes, 
which decay exponentially.

The important information in this method is that $w$ is a linear combination of only unstable modes.
To find the coefficients of this linear combination using the method given here, we need to know all the CLVs and $v^\infty$.
This method is infeasible since the computational cost will be high to find all CLVs and $v^\infty$ is unknown \textit{a priori}.
These difficulties are overcome in the next subsection.

\subsection{Computing \texorpdfstring{$v^\perp$}{vperp} by NILSS}\label{ss:w by wj}

In NILSS, we compute $v=v^* + w$ such that $v^{\perp}\approx v^{\infty\perp}$.
More specifically, we want the integration of $v^{\perp}$ to approximate $v^{\infty\perp}$ so that later, when computing the sensitivity via equation (\ref{eq:derivative}), $v^{\perp}$ yields a result close to the result given by $v^{\infty\perp}$.
To achieve this, we solve the NILSS problem on a single time segment, which is to minimize the $L^2$ norm of $v^{\perp}=v^{*\perp} + W^\perp a$:
\begin{equation} \label{eq:nilssprb}
  \min_a \frac12\integrate ( v^{*\perp} + W^\perp a)^{T} ( v^{*\perp} + W^\perp a) \; dt ,
\end{equation}
which is simply a least squares problem with arguments $ a \in \R^M $, where $M$ is an integer larger than the number of positive LEs $\mus$.
Here $v^*$ is the conventional tangent solution, 
and $W^\perp(t)$ is a matrix whose columns are homogeneous tangent solutions $\{w_j^\perp(t), j = 1,...,M\}$.
The initial conditions $ \{w_j(t=0) \}$ are randomized unit vectors in $ \R^m $.

First we need to see that a desired $v^\perp$ exists in the feasible solution space, 
or that some $a$ can yield a desired $v^\perp=v^{*\perp} + W^\perp a$. 
Our discussion in the last subsection confirms the existence if we use unstable CLVs instead of $W$.
Moreover, \cite{Benettin1980_LE} proves that as time evolves, the span of $\{w_j^\perp(t), j = 1,...,M\}$ converges to the span of the CLVs $\{\zeta_j^\perp(t), j = 1,...,M\}$ with the largest $M$ LEs.
As a result, replacing unstable CLVs by $W^\perp$ gives a feasible solution space that contains a $v^\perp$ such that $v^{\perp}\approx v^{\infty\perp}$.

\begin{figure}[htb]
  \centering
  \includegraphics[width=0.5\textwidth]{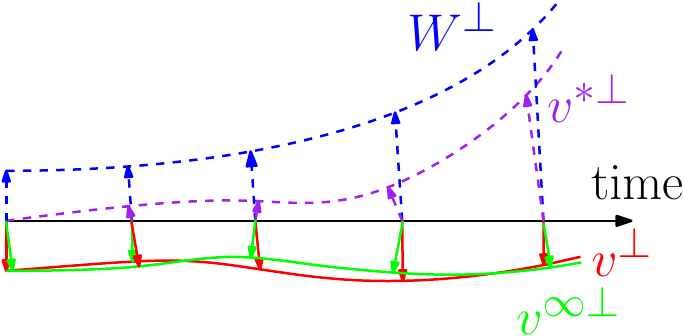}
  \caption{Intuition of NILSS: through minimization over $\|v^\perp\|$, we find a column vector $a$,
  such that $v^\perp =v^{*\perp} + W^\perp a \approx v^{\infty\perp}$.
  This is because most unstable components in $v^{*\perp}- v^{\infty\perp}$ are subtracted by $ W^\perp a $ during the minimization.}
  \label{f:v_vstar_w}
\end{figure}

Next, we need to rationalize that minimizing the $L^2$ norm of $v^{\perp}=v^{*\perp} + W^\perp a$ yields $v^{\perp}\approx v^{\infty\perp}$.
First, we notice that $v^{\perp}$ can be written as the summation of $v^{\infty\perp}$ and some homogeneous tangents.
Because $v^{\infty\perp}$ is bounded and the unstable modes (now approximated by the span of $W$) grow exponentially, then minimizing $v^{\perp}$ over a long trajectory implies the difference $v^{\perp}-v^{\infty\perp}$ cannot contain significant unstable components. 
Although stable modes may be left in this difference, they decay exponentially.
The effect of the minimization is illustrated by figure \ref{f:v_vstar_w}.

\subsection{Computing \texorpdfstring{$d{\avg{J}_\infty}/ds$}{dJds} from the tangent solution}

\label{ss:compute djds1}
Since $v-v^\perp$ is parallel to $f$, we can define $\xi$ as the scalar which satisfies: 
\begin{equation} \label{eq:xi}
\xi f = v-v^\perp .
\end{equation}
To find a pair $(v^\perp, \xi)$, we first find a $v$ which solves equation (\ref{eq:tangent equation}), 
project $v$ onto the subspace $V^\perp$ to find $v^\perp$, 
then use equation (\ref{eq:xi}) to find $\xi$.

Once we obtain the solution vector $a$ of the NILSS problem, we can construct $v=v^*+Wa$ and compute the corresponding $\xi$.
Then we have the following approximation for $d{\avg{J}_\infty}/ds$:
\begin{equation} \label{eq:derivative}
	\dd{\avg{J}_\infty}{s} \approx 
	\frac 1T \left[
	\int_{0}^{T} \left(\partial_u J \, v+ \partial_sJ \right)dt 
	+\left. \xi\right\vert^T_0 \avg{J}_T
	-\left. \left(\xi J \right) \right\vert^T_0\right] ,
\end{equation}
where $\avg{J}_T$ is defined in equation (\ref{eq:average J}). 
Notice that in equation (\ref{eq:derivative}), we use the tangent solution $v$ instead of its projection $v^\perp$.
The derivation of equation (\ref{eq:derivative}) is in \ref{ss:derive djds}.

\subsection{Benefits of NILSS}

In NILSS, the optimization problem is comprised of only a small part of the computational cost, since there are only $M$ arguments in equation (\ref{eq:nilssprb}). 
The main cost comes from setting-up the optimization problem by computing $v^{*\perp}$ and $w_1^\perp(t),..., w_M^\perp(t)$. 
Hence the cost of NILSS is proportional to the number of unstable modes $\mus$.
For engineering problems, $\mus$ is usually much smaller than $m$; thus, the cost of NILSS is low.

NILSS is easily implemented with existing tangent solvers.
The data used in the NILSS problem are $v^{*\perp}$ and $\{w_j^\perp\}$.
Here $v^*$ is the result of a conventional tangent solver.
$\{w_j\}$ are given by homogeneous tangent solvers, 
which can be obtained by setting the right hand side in equation (\ref{eq:tangent equation}) to zero in conventional tangent solvers.
Once we have $v^*$ and $\{w_j\}$, $v^{*\perp}$ and $\{w_j^\perp\}$ can be computed by orthogonal projection onto $V^\perp(u)$,
as shown in equation (\ref{eq:vperp projection}).

A beneficial side-effect is that NILSS uses less computer memory than LSS.
Furthermore, the tangent solutions used in NILSS do not need to be saved in the computer memory concurrently.
NILSS can use tangent solutions saved on an external hard drive, which can then be read in pairs to compute their inner product;
this may reduce the computational speed, but further saves computer memory.

Another way to compute those tangent solutions is to approximate them by finite difference solutions.
This leads to the finite difference NILSS (FD-NILSS)  \cite{Ni_fdNILSS}.
In this way, FD-NILSS requires only primal simulation and no longer the tangent solvers.

\section{Another derivation of the non-intrusive formulation} \label{s:short derivation from LSS}

\noindent{}
\nax{
  (This section is newly added, and we appreciate feedback from readers.)
}

In this sections we offer another derivation of the non-intrusive formulation,
based on reducing the feasible set in the least squares shadowing (LSS) algorithm.
This derivation also requires several definitions given in section~\ref{s:longer derivation of nilss}, briefly summarized below.
We define inhomogeneous tangent solution as $v=\partial u/\partial s$, governed by equation~\eqref{eq:tangent equation}.
We define the conventional tangent solution, $v^*$, as the inhomogeneous tangent solution with zero initial condition.
We define homogeneous tangent solution as $w=\partial u/\partial \phi$, governed by equation~\eqref{eq:homogeneous tangent equation}.
We define CLVs as homogeneous tangent solutions whose norms grow as exponential functions of time.

The butterfly effect states that, most perturbations on initial conditions 
lead to new trajectories diverging quickly from the original trajectory.
The linearized version of this statement is, for most initial conditions,
homogeneous tangent solutions grow exponentially fast.
Similarly, for most perturbations on the parameter $s$,
trajectories starting from the same initial condition also diverge quickly from original.
The linearized version of this statement is, for most $\partial_s f$, $v^*$ grows exponentially fast.

However, the shadowing lemma states that, if perturb both the initial condition and parameter, and coordinate the two perturbations carefully,
we can find a new trajectory such that it always lies close to the original trajectory \cite{Bowen_shadowing}.
The linearized version of this statement is, if we carefully choose initial conditions of inhomogeneous tangent solutions,
we can find a $v^\infty$, called shadowing direction, such that its perpendicular component is uniformly bounded.
This bounded property justifies the interchange of limits in equation~\eqref{eq:commutelimit},
and we can now use average of the perturbation, described by tangent solutions, to compute perturbation of the averaged objective.

LSS states that, the boundedness of $v^{\infty\perp}$ can be mimicked by minimizing its integrated $L^2$ norm.
That is, we can perform a minimization within all inhomogeneous tangent solutions, to approximate the shadowing direction on a finite trajectory:
\begin{equation} \begin{split} \label{e:lss}
  &\min_{v} \frac 12 \integrate \|v^\perp\|^2,
  \quad  \mbox{s.t. }
  \dd{v}{t} - \partial_u f v = \partial_s f.
\end{split}\end{equation}
Notice that our version of LSS in this paper is different from its original form given by Wang \cite{Wang_ODE_LSS}:
we introduce the perpendicular projection operator to help separate the computation of tangent solutions, 
computation of time dilation term, and the minimization.
These separations are important for developing the non-intrusive formulation.

Instead of solving the LSS problem by the KKT conditions,
we can think of starting from a particular solution inside the feasible set of all inhomogeneous tangent solutions,
say $v^{*\perp}$, and see how to modify this starting point to approach the desired shadowing direction, $v^{\infty\perp}$.
The difference $v^{*\perp}-v^{\infty\perp}$ is homogeneous tangent solution;
moreover, since the neutral CLV is projected out by the perpendicular projection,
this difference can be decomposed to stable and unstable CLVs.
Since stable CLVs decays and unstable grows exponentially, 
we only need to subtract from $v^{*\perp}$ the unstable parts of this homogeneous difference,
to obtain a good approximation of shadowing direction.

In other words, instead of searching within the large feasible set of all inhomogeneous tangent solutions,
we only need to search those can be written as $v=v^*+\sum_{j=1}^{m_{us}} a_j \zeta_j(u)$,
where $\{\zeta_j\}_{j=1}^{m_{us}}$ are all unstable CLVs.
Together with the fact that the span of first several CLVs is approximated, in the long time,
by the span of same number of randomly initiated homogeneous tangent solutions \cite{Benettin1980_LE}, 
we obtain the non-intrusive formulation of LSS:
\begin{equation} \begin{split} 
  &\min_{a} \frac 12 \integrate \|v^\perp\|^2,
  \quad  \mbox{s.t. }
  v =  v^* + Wa \,,
\end{split}\end{equation}
where $a\in \R^M$, and $W$ is a time-dependent matrix of homogeneous tangent solutions.
We typically take $M$ slightly larger than $\mus$,
to guarantee that the unstable subspace is approximately included in the span of $W$ after a finite time.
Notice that we reparameterize the minimization problem,
changing the arguments from $v$ to $a$, coefficients of homogeneous tangent solutions.
With this $v$, we can then compute sensitivities as discussed in section~\ref{ss:compute djds1}.

To conclude, the non-intrusive formulation reduces the feasible set in LSS to a smaller set affine to the unstable subspace.
For many engineering problems, $\mus$ is much smaller than $m$; thus, the cost of NILSS is much lower than LSS.
This new parameterization also separates the computation of tangent solutions from minimization,
hence NILSS requires only minor modifications on existing tangent solvers.
The name `non-intrusive' comes from how we derived the method by reparametrization,
the benefit on easy implementation, and the coincidence with the first author's last name.

\section{Numerical aspects of NILSS} \label{s:Tangent NILSS Algorithm}

In this section, we first address the numerical stability of the algorithm by rescaling 
$ v^{*\perp} $ and $ W^\perp $ after every short segment of time $\Delta T$. 
Then, we discuss the criterion for determining the number of homogeneous solutions $M$ and segment length $ \Delta T $.

\subsection{Solving NILSS on multiple time segments}

Since both $ v^{*\perp} $ and $ W^\perp $ grow exponentially, 
the round-off error when storing them in the computer become non-negligible over time.
The growth in $ v^{*\perp} $ and $ W^\perp $ will also generate an ill-conditioned covariance matrix $ (W^\perp)^T  W^\perp$, 
since all $\{w_j^\perp\}$ will eventually be dominated by the fastest growing unstable CLV.
This subsection shows how to prevent this by partitioning a long trajectory into a series of shorter segments:
this idea is similar to a method for computing LE \cite{Benettin1980_LE}, 
but we normalize not only homogeneous tangent solutions, but also inhomogeneous solutions.

\begin{figure}[htb]
  \centering
  \includegraphics[width=0.5\textwidth]{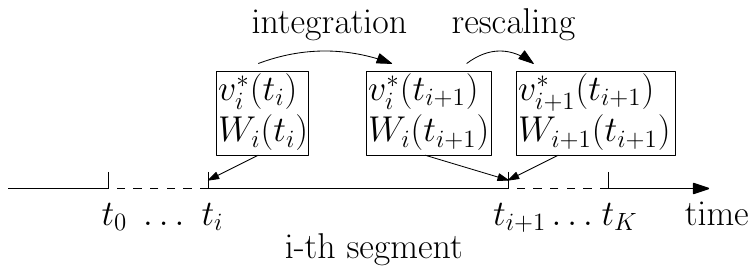}
  \caption{Notations used for NILSS, $ t_0 = 0, t_K  = T $}
  \label{f:subscriptExplain}
\end{figure}

We partition the time domain into $K$ time segments $[t_0,t_1], [t_1,t_2],\ldots,[t_{K-1},t_K]$, with $ t_0=0, t_K=T $. 
Next, we define time segment $ i $ as $ [t_i,t_{i+1}], i=0,\ldots,K-1 $.
For each time segment $ i $, we define an inhomogeneous solution $ \{v^{*} _i\} $
and homogeneous solutions $ \{W_i\} $, such that each $ W_i = [w_{i1},\cdots,w_{iM}] $.
This notation is depicted in figure \ref{f:subscriptExplain}.

We want to rescale and orthogonalize $ v_i^{*\perp} $ and $W_i^\perp $ 
at the end of each segment so that they do not grow too large or become dominated by the fastest growing CLV.
We also want to keep the affine vector space $ v^{*\perp}  + span(W^\perp) $ the same across interfaces between contingent segments, so that we can recover a continuous $v^\perp$:
\begin{equation} \label{eq:conti space}
  v^{*\perp} _{i}(t_i) + span\left(W^\perp_{i}(t_i)\right)
  = v^{*\perp} _{i-1}(t_i) + span\left(W^\perp_{i-1}(t_i)\right) ,
\end{equation}
where $ span(W^\perp) $ is the vector space spanned by the column vectors of $W^\perp $.

To achieve this, we first orthonormalize $W^\perp$ via a QR decomposition:
\begin{equation} \label{eq:qr}
  W^\perp_i(t_{i+1}) = Q_{i+1} R_{i+1} .
\end{equation}
We set the initial conditions of the next tangent segment to
\begin{equation} \label{eq:w initial}
  W_{i+1}(t_{i+1}) = Q_{i+1} .
\end{equation}
In QR factorization, column spaces of $Q_i$ and $W^\perp_i$ are equal if the column vectors in $W^\perp_i$ are linearly independent.
Indeed, the linear independence of the initial condition of $W^\perp_i$ can be preserved after $\Delta T$, if $f$ is Lipschitz continuous.
Hence, $ span\left(W^\perp_{i+1}(t_{i+1})\right) \;=\; span\left(W^\perp_{i}(t_{i+1})\right)$,
where $W^\perp_{i}(t_{i+1})=Q_{i+1}$.

We subtract from $v^{*\perp}$ its orthogonal projection on $W^\perp$ to obtain the initial condition of the next time segment:
\begin{equation}\label{eq:v* initial}
  v^{*} _{i+1}(t_{i+1}) = v^{*\perp} _i(t_{i+1}) - Q_{i+1} b_{i+1} , 
\end{equation}
where $ b_{i+1} = Q_{i+1}^T v^{*\perp} _i(t_{i+1}) $.
$ v^{*\perp}_{i+1}(t_{i+1})$ is still in the affine space $v^{*\perp}_{i}(t_{i+1}) + span\left(W^\perp_{i}(t_{i+1})\right)$.
The norm of $v^{*\perp}_{i}(t_i)$ is reduced, since the unstable modes in it are subtracted through the projection.

We want to recover a continuous $ v^{\perp} $ over the whole trajectory,
which is dissected to $v_i$ on the $i$-th segment,
which is further represented by $v_i=v^*_i+W_ia_i$.
Now equation~\eqref{eq:conti space} is satisfied, for any $a_{i}$, there exists $a_{i+1}$ such that:
\begin{equation}
  v^{*\perp} _{i+1}(t_{i+1}) + Q_{i+1} a_{i+1}
  = v^{*\perp} _{i}(t_{i+1}) + W^\perp_{i}(t_{i+1}) a_{i}.
\end{equation}
This allows the continuity condition at time $t_{i+1}$:
\begin{equation} \label{eq:conti vperp}
  v^{\perp}_{i+1}(t_{i+1}) \;=\; v^{\perp}_{i}(t_{i+1}) .
\end{equation}
By applying equation~\eqref{eq:qr}, \eqref{eq:w initial}, and \eqref{eq:v* initial}, we can show this is equivalent to:
\begin{equation} \label{eq:cts requirement}
  a_{i+1} = R_{i+1} a_{i} + b_{i+1} .
\end{equation}
This continuity condition assures the solution $v^\perp$ over multiple time segments is equivalent to that over a longer segment.
However, rescaling $ v^{*\perp} $ and $ W^\perp $ at the end of each time segment prevents them from growing too large.

\subsection{Determining parameters for NILSS}\label{ss:determine parameters}

There are two parameters in NILSS that users should choose: 
the number of homogeneous solutions $M$ and the length of each time segment $\Delta T$.
This subsection discusses the criteria for determining these parameters.
Once the parameters are determined, we can proceed to following subsections about the detailed algorithm of NILSS.

$M$ is determined based on the Lyapunov Exponents (LE), which are byproducts of NILSS.
According to \cite{Benettin1980_LE}, $ \lambda_j $, the j-th largest LE, is computed by:
\begin{equation}
  \lambda_j \approx \frac{1}{K \Delta T}\sum_{i = 1}^{K} \log(\left|d_{ij} \right|),
\end{equation}
where $ d_{ij} $ is the j-th diagonal element in $R_i$.
Notice that the computation of $\{R_i\}$ only require $W$ but not $v^*$.
As we shall see in the detailed algorithm later, NILSS can compute homogeneous solutions $W$ before $v^*$. 
At the stage of computing $W$, we can gradually increase $M$ and compute more LEs, which appear in decreasing order.
Once we have a negative LE, we know that we have found all positive LEs.

$\Delta T$ is determined by the constraint that the CLV with the largest LE does not dominate the $M$-th CLV.
If we assume the largest LE is $ \lambda_{1}$ and the $M$-th LE is $ \lambda_{M}$, 
then the ratio between the norm of these two CLVs satisfies:
\begin{equation}
  \frac{\|\zeta_1^\perp (u(t))\| / \|\zeta_1^\perp (u(0))\|}
  {\|\zeta_M^\perp (u(t))\| / \|\zeta_M^\perp (u(0))\|} \approx \exp((\lambda_1-\lambda_M)t).
\end{equation}
This suggests that the ratio between the fastest growing and the M-th CLV grow about three times larger after a time span $(\lambda_{1}-\lambda_{M})^{-1}$. 
If $\Delta T$ is large, the covariance matrix $C_i$ in equation (\ref{eq:covariant matrix}) will be ill-conditioned, 
which could pose a numerical problem.
To prevent this from happening, we rescale $ W^\perp $ and $ v^{*\perp} $ after $  \Delta T \lessapprox (\lambda_{1}-\lambda_{M})^{-1}$.
On the other hand, when $ \Delta T$ get smaller, there are more segments,
which leads to a larger optimization problem in equation (\ref{eq:NILSS on multiple segments}).
This concern on cost gives the lower bound of $\Delta T$.

\section{Procedure list of NILSS}

In this section we provide a walk-through of the NILSS algorithm.

\subsection{Pre-processing}

First, we integrate equation (\ref{eq:dynamical system}) over a sufficient period before $ t=0 $ so that $ u $ is on the attractor at the beginning of our algorithm. 
Then, we integrate equation (\ref{eq:dynamical system}) from $t=0$ to $t=T$ to obtain the primal solution $ u(t) $.

\subsection{Computing the homogeneous solution \texorpdfstring{$\{W_i\}$} {Wi} }

We compute one inhomogeneous and $M$ homogeneous tangent equations for each of the $K$ time segments
$[t_0,t_1]$,$\ldots$,$[t_{K-1},t_K]$, where $ t_0 = 0, t_K = \tend $. 
Time segment $ i $ is with the range $ [t_{i}, t_{i+1}] $.
This notation is the same as those found in figure \ref{f:subscriptExplain}.

We start at the first segment with random initial conditions for each column vector in $W$:
\begin{equation} \begin{split}
  W_{0}(0) &= [w_{01}(0), \ldots, w_{0M}(0)], \quad \text{with } w_{0j}(0)\in V^\perp(u(0)) .
\end{split} \end{equation}

Then, we proceed with the following algorithm, which starts at $i=0$.
\begin{enumerate}
  \item
  For each $j=1,\cdots, M$, we start from the initial conditions $ \{w_{ij}(t_{i})\} $.
  We then integrate equation (\ref{eq:homogeneous tangent equation}) to obtain $w_{ij}(t)$,$t\in [t_{i},t_{i+1}]$.
  We compute the orthogonal projection onto $V^\perp$ using equation (\ref{eq:vperp projection}):
  \begin{equation}
  W_i^\perp(t) = [w_{i1}^\perp(t), \ldots, w_{iM}^\perp(t)], \quad t\in [t_{i},t_{i+1}] .
  \end{equation}

  \item 
  Then, we compute and store the
  \begin{equation} \label{eq:covariant matrix}
  C_i = \int_{t_{i}}^{t_{i+1}} (W_i^\perp)^T W_i^\perp dt . 
  \end{equation}
  
  \item 
  We orthonormalize $W_i^\perp(t_{i+1})$ with a QR decomposition under the Euclidean norm:
  \begin{equation}
    W_i^\perp(t_{i+1}) = Q_{i+1} R_{i+1}. 
  \end{equation}
  Then, we store $R_i$ and set the initial conditions of the next segment to
  \begin{equation} 
    W_{i+1}(t_{i+1}) = Q_{i+1}. 
  \end{equation}
  
  \item 
  Finally, we let $i = i+1$, after which we go to Step 1 unless $ i=K $, in which case we proceed to section \ref{ss:compute vstar}.
\end{enumerate}

Here we compute $ \{w_{ij}\} $ from equation (\ref{eq:homogeneous tangent equation}). 
They may also be computed as the difference between two inhomogeneous tangent solutions:
\begin{equation}
  \{w_{ij}\}  =v ^w_{ij} -v^0_i,
\end{equation}
where $v ^w_{ij}$ has same initial condition as $w_{ij}$ and $ v^0_i $ has a zero initial condition at $t_i$.
This way of computing homogeneous tangents no longer requires a separate homogeneous tangent solver.

\subsection{Computing the inhomogeneous solution \texorpdfstring{$\{v^*_i\}$} {vstar i} } 
\label{ss:compute vstar}

We start at the first time segment with initial condition: $v^*_{0}(0) = 0$, then proceed with the following algorithm starting at $i=0$.

\begin{enumerate}
  \item 
  Starting from the initial condition $v^*_i(t_{i})$,
  integrate the inhomogeneous equation (\ref{eq:tangent equation}) to obtain $v^*_i(t)$, $t\in [t_{i},t_{i+1}]$.
  Through equation (\ref{eq:vperp projection}), 
  we compute the orthogonal projection $v_i^{*\perp}(t)$, $t\in [t_{i},t_{i+1}]$.
  
  \item 
  Compute and store
  \begin{equation} 
    d_i = \int_{t_{i}}^{t_{i+1}} {W_i^\perp}^T v^{*\perp}_i dt .
  \end{equation}
  
  \item 
  Orthogonalize $v^{*\perp}_i(t_{i+1})$ with respect to $W^{\perp}_{i+1}(t_{i+1})$ to obtain the initial condition of the next time segment:
  \begin{equation}
    v^*_{i+1}(t_{i+1}) = v^{*\perp}_i(t_{i+1}) - Q_{i+1} b_{i+1} ,
  \end{equation}
  where
  \begin{equation} 
    b_{i+1} = Q_{i+1}^T v^{*\perp}_i(t_{i+1}) 
  \end{equation}
  should be stored.
  
  \item 
  Let $i = i+1$.  Go to Step 1 unless $ i=K $, in which case we proceed to section \ref{ss:compute v}.
  
\end{enumerate}

Here we compute the inhomogeneous solution $ v^*_i $ and homogeneous solution $ W_i $ separately.
By doing this, we can first find all positive LEs by gradually increasing $M$, since the computation of LE only requires homogeneous solutions.
Once $M$ is determined, we can go on to compute $v^*$.
If we already know the number of positive LEs, then $ v^*_i $ and $ W_i $ can be computed simultaneously.

\subsection{Computing \texorpdfstring{$v$}{v} } \label{ss:compute v}

Here we compute $\{v_i\}$ for each segment, with $v^{\perp}_i$ continuous across different segments.
The minimization in equation (\ref{eq:nilssprb}) becomes:
\begin{equation} 
  \sum_{i=0}^{K-1} \int_{t_{i}}^{t_{i+1}} \left[
  (v_i^{*\perp})^T v_i^{*\perp} + 2 (v_i^{*\perp})^T W^\perp_i a_i + a_i^T (W_i^\perp)^T W_i^\perp a_i\right] dt ,
\end{equation}
where $\{a_i\in\R^{M},i=0,\ldots,K-1\}$. 
Other than a constant contribution from $(v_i^{*\perp})^T v_i^{*\perp}$, which is independent of $\{a_i\}$, 
we should choose $\{a_i\}$ via a least squares problem.
Combining with the continuity constraints in equation (\ref{eq:cts requirement}), we obtain the NILSS problem for multiple time segments:
\begin{equation} \begin{split}\label{eq:NILSS on multiple segments}
  &\min_{\{a_i\}} \sum_{i=0}^{K-1} 2 d_i^T a_i
  + a_i^T C_i a_i \\
  \mbox{s.t. }& 
  a_{i} = R_{i} a_{i-1} + b_{i}
  \quad i=1,\ldots,K-1.
\end{split}\end{equation}
Once $ \{a_i\} $ is obtained, we can compute $ v_i $ within each time segment $t\in[t_i,t_{i+1}]$ via the expression
\begin{equation} \label{eq:v_i}
 v_i(t) = v^*_i(t) + W_i(t) a_i .
\end{equation}

\subsection{Computing \texorpdfstring{$ \xi_i $}{xi i}}

For each segment $i$, we define $\xi_{i}(t)$ by plugging $v$ into equation (\ref{eq:xi}) to arrive at
\begin{equation}\label{eq:xi_i}
\xi_i f = v_i-v_i^{\perp} .
\end{equation}

In fact, we only need to know the value of $ \xi_i $ at the beginning and end of each segment, that is:
\begin{equation} \label{eq:xi end value}
\begin{split}
\xi_i(t_i) &= 0 \;;\\
\xi_i(t_{i+1}) &= \frac{( v_i(t_{i+1}) )^T f(u(t_{i+1}))}{f(u(t_{i+1}))^T f(u(t_{i+1}))} \;.
\end{split}
\end{equation}
In equation \ref{eq:xi end value}, we used the fact that at the beginning of each segment, $v^*_i$ and $W_i$ are in $V^\perp$, hence so is $v_i$.

On each segment $i$, here we first use a linear combination of  $v^*_i$ and $\{w_{ij}, j = 1, \cdots, M\}$ to compute $v_i$, as done in equation (\ref{eq:v_i}), then use $v_i$ compute $\xi_i$. 
Alternatively, we can first compute the contribution of $v^*_i$ and $\{w_{ij}, j = 1, \cdots, M\}$ in $\xi_i$, and then compute $\xi_i$ through a linear combination with the same coefficient vector $a_i$ as in equation (\ref{eq:v_i}).

\subsection{Computing \texorpdfstring{$ d\avg{J}_\infty/ds $}{dJds} } 
\label{ss:compute djds}

Once $ v(t) $ is obtained, $ d\avg{J}_\infty / ds $ is computed via
\begin{equation} \label{eq:dJds seg}
  \frac 1T\sum_{i=0}^{K-1} \left[
  \int_{t_i}^{t_{i+1}} 
  \left(\partial_u J \, v_i+ \partial_sJ \right)dt 
  + \xi_i(t_{i+1}) (\avg{J}_T - J(t_{i+1})) \right] .
\end{equation}
The derivation of equation (\ref{eq:dJds seg}) from equation (\ref{eq:derivative}) is in \ref{ss:derive djds multiple segments}.

Alternatively, the sensitivity can be computed without explicitly determining $\{v_i(t)\}$.  
The sensitivity contribution of each $v_i(t)$ can be computed from a linear combination of the contributions of $v^*_i$ and $w_{ij}$, 
with $a_i$ being the coefficients.

\section{Applications on chaotic partial differential equation systems}

\subsection{Numerical Results on Lorenz attractor}

We apply NILSS to the Lorenz 63 system. 
\footnote{A python code implementing NILSS is available at https://github.com/niangxiu/nilss.
This code is not optimized in performance, because we want to keep it short for easy understanding.}
There are three states $u = [x, y, z]$, so $ m=3 $.
The governing equation is:
\begin{equation}
  \dd{x}{t} = \sigma(y-x), 
  \quad \dd{y}{t}=x(\rho-z)-y,
  \quad \dd{z}{t} = xy-\beta z .
\end{equation}
In our current numerical example, $ \sigma = 10, \beta = 8 / 3 $.

The parameter of the system is $ \rho $, which varies in range $ [2,45] $.
The Lorenz 63 system has different behaviors when $ \rho $ changes \cite{sparrow2012lorenz}:
\begin{itemize}
  \item $ 2\le\rho <24.7 $, two fixed-point attractors. 
  \item $ 24.7\le\rho <31 $, one quasi-hyperbolic strange attractor.
  \item $ 31 \le \rho \le 45 $, one non-hyperbolic attractor.
\end{itemize}
In none of these cases the dynamical system strictly satisfies our assumptions 
that there exists a full set of CLVs for all states on the attractor;
however, as we shall see, NILSS still gives meaningful results.
The instantaneous objective function is $J(u) = z$, so the objective is:
\begin{equation}
  \avg{J}_\infty = \lim\limits_{T\rightarrow \infty} \frac1T \integrate z\, dt \,.
\end{equation}
We use $\avg{J}_{T'}$ to approximate $\avg{J}_\infty$, where $T'=$ 500 time units.
Moreover, the initial state $u_0$ of each $\rho$ is randomized.

When solving the primal solution $ u = (x,y,z)^T$, we use RK-4 with time step size $0.01$. 
Each segment has 200 steps, or 2 time units.
We perform NILSS over $ K =50 $ segments, i.e., $T=$ 100 time units.

The LEs of the Lorenz 63 system should satisfy the following constraints \cite{bovy2004lyapunov}:
\begin{equation}
\begin{split}
\lambda_1 + \lambda_2 + \lambda_3 &= -(1+\sigma+\beta) ;\\
\lambda_3 &= 0 .
\end{split}
\end{equation}
Here $\lambda_3$ is the LE whose corresponding CLV is parallel to $du/dt$.
Since $\lambda_1+ \lambda_2<0$, there are at most 1 positive LE.
Hence we set the number of homogeneous solutions to be $M=1$.

\begin{figure}[htb]
  \centering
  \includegraphics[width=0.49\textwidth]{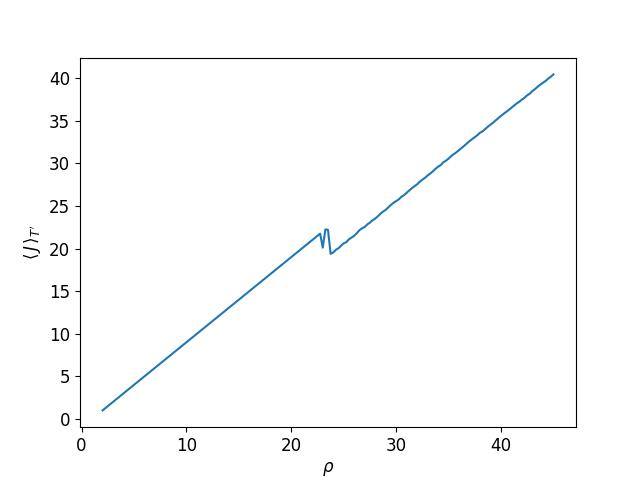}
  \includegraphics[width=0.49\textwidth]{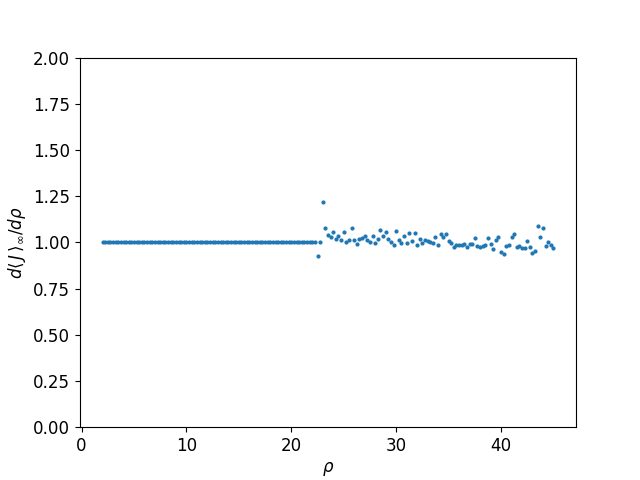}
  \caption{Application on Lorenz system.
    Left: Averaged objective versus parameter for the Lorenz 63 system, with $ \sigma = 10, \beta = 8 / 3 $, $T'=500$ time units.
    Right: $ d\avg{J}_\infty/d\rho$ computed for each $ \rho $ via NILSS. 
    Length of trajectory is $ T=100 $, which is partitioned into 50 segments of $\Delta T=2$. 
    NILSS uses one homogeneous tangent solution.}
  \label{f:Lorenz}
\end{figure}

With above setting, we compute $\avg{J}_\infty $ and $ d\avg{J}_\infty/d\rho $.
The results are shown in figure \ref{f:Lorenz}.
The flaw shown in the left is also observed in other numerical results such as those found in \cite{Lea2000}.
This flaw corresponds to the onset of chaos around $\rho=24.7$.
For smaller $\rho$, the system has two fixed-points, and the sensitivity results, via NILSS, show no oscillation.
When the system develops into chaos, the sensitivity results begin to oscillate because, on a finite trajectory, they depend on the random-valued initial conditions $u_0$ and $W_0(0)$.
Nevertheless, figure \ref{f:Lorenz} shows that the true value of $ d\avg{J}_\infty/d{\rho} $ is approximately 1 for all $ \rho $.
The sensitivities computed with NILSS agree with this observation.

\subsection{Numerical Results on CFD Simulation of flow over a backward-facing step}

\begin{figure}[htb]
  \centering
  \includegraphics[trim={0 1cm 0cm 0cm}, clip=true, width=0.8\textwidth]{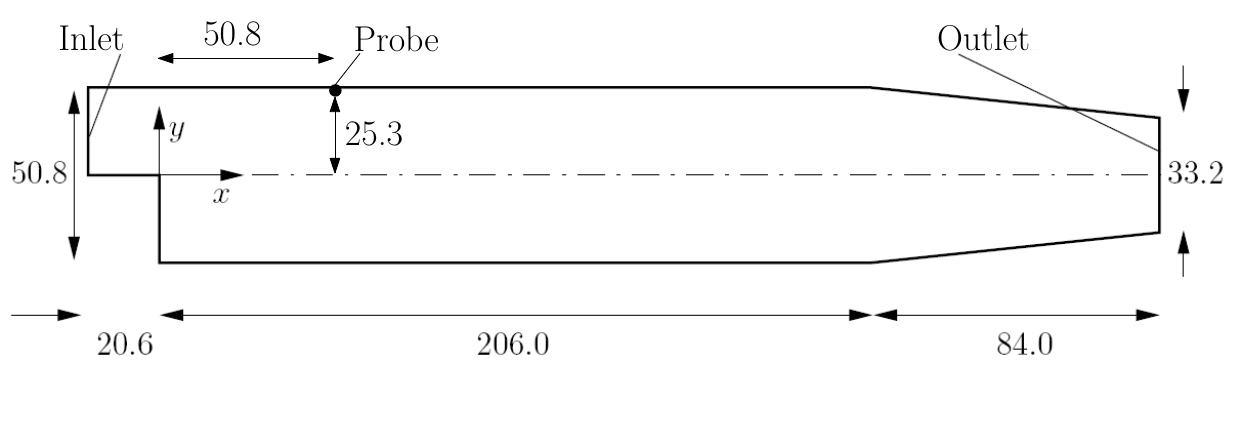}
  \caption{Geometry used in the simulation of a chaotic flow over a backward-facing step, dimensions in mm.
  All boundaries except inlet/outlet are solid walls.}
  \label{f:pitzdailyGeometry}
\end{figure}

We apply NILSS to a chaotic flow over a backward-facing step.
Specifically, we use the same geometry and mesh as in the PitzDaily tutorial of OpenFOAM 4.0, which is modeled from the experiment by Pitz and Daily \cite{Pitz1983}.
This problem is a two-dimensional flow over a backward-facing step near the inlet and a contracting nozzle at the outlet.
The geometry is shown in figure \ref{f:pitzdailyGeometry}.

\begin{figure}[htb]
  \centering
  \includegraphics[trim={0 15cm 0 15cm},clip, width =0.8\textwidth]{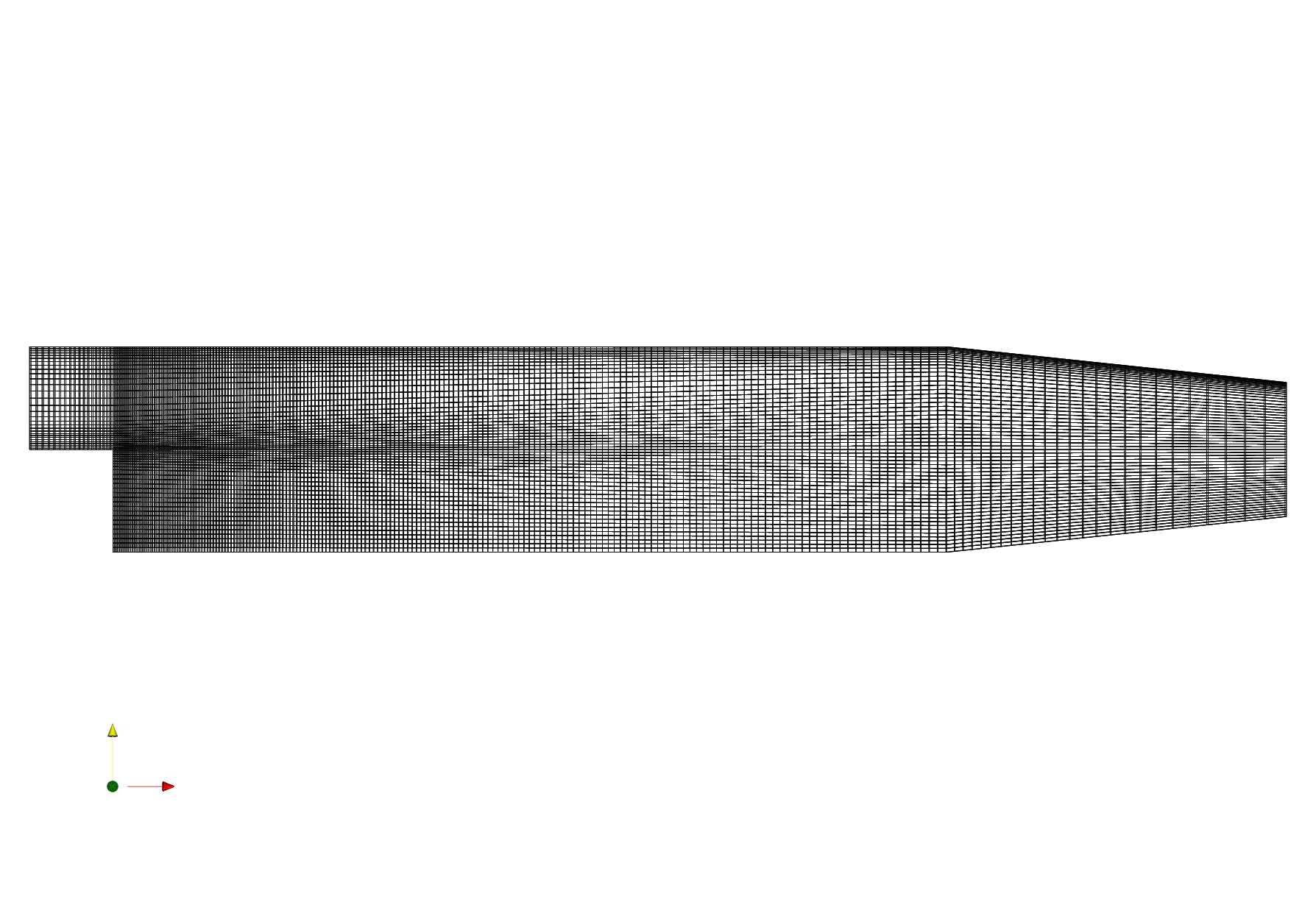}
  \caption{Mesh of test case, as provided in the tutorial of OpenFOAM 4.0 }
  \label{f:pitzdailyMesh}
\end{figure}

For the numerical simulation, we use OpenFOAM 4.0 as the solver.
We use the mesh provided in the tutorial: there are 12225 cells, as shown in figure \ref{f:pitzdailyMesh}.
We solve the incompressible Navier-Stokes equation via pisoFOAM.
We use the second-order finite volume scheme and the time-integration method is PISO (Pressure Implicit with Splitting of Operator) with a time step size $1\times 10^{-5}$ second.
We use dynamic one equation eddy-viscosity model as turbulence model \cite{kim1995new}.
The viscosity is $1\times 10^{-5}m^2/s$.

We set no-slip wall conditions for all boundaries except for the inlet and outlet.
The velocity at the inlet boundary takes a uniform fixed value in the x-direction, the norm of which is the parameter of this problem.
For the base case, we set the inlet velocity to $ U = (10,0,0)m/s $. 
For the outlet, we use the `inletOutlet' option, which is to switch between the zero value and the zero gradient boundary condition, 
depending on the flow direction.

\begin{figure}[htb]
  \centering
  \includegraphics[trim={0 3.5cm 0 3cm},clip, width = 0.8\textwidth]{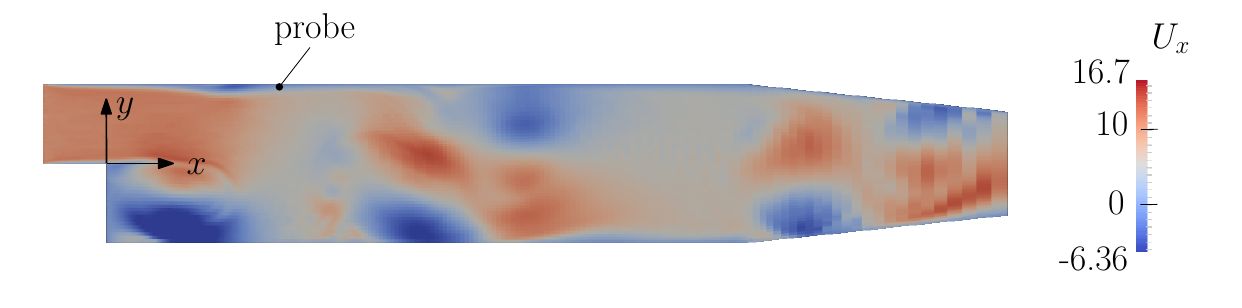}
  \caption{Flow field at time 0.091. Plotted by x-directional velocity $U_x$.}
  \label{f:pitzdailyField}
\end{figure}

With the above settings, a typical snapshot of the flow field is shown in figure \ref{f:pitzdailyField}.
The flow is chaotic but not turbulent, since it is two-dimensional.
Moreover, for a real-life problem, like the current one, 
there is no guarantee that all of our assumptions made when developing NILSS will be satisfied.
However, as we shall see, NILSS still gives meaningful results.

The parameter in this problem is the x-directional velocity at the inlet, $U_{x0}$.
We use four different objectives: the long-time average of $U_x/10$, $(U_x/10)^2$, $(U_x/10)^4$, and $(U_x/10)^8$, where $U_x$ is the x-direction velocity at a probe at coordinate (50.8 mm, 25.3 mm). 
The location of the probe is very close to the upper surface, as shown in figure \ref{f:pitzdailyField}.

\begin{figure}[!htb]
	\centering
	\begin{subfloat}
		{\includegraphics[trim=0cm 0cm 1cm 0.5cm, clip=true, width=0.4\textwidth]{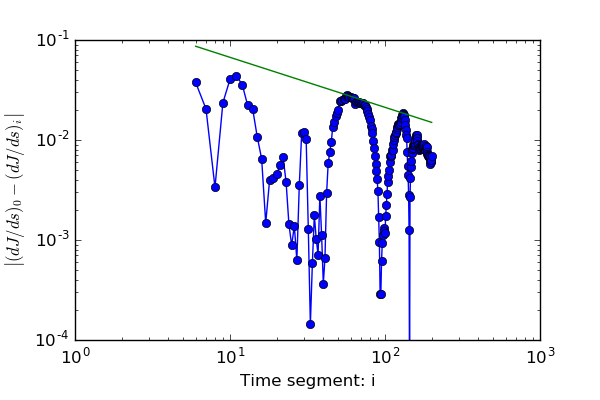}}
		{\includegraphics[trim=0cm 0cm 1cm 0.5cm, clip=true, width=0.4\textwidth]{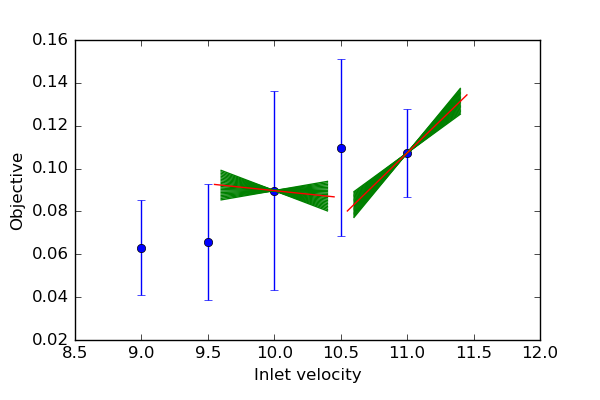}}
	\end{subfloat}\\
	\begin{subfloat}
		{\includegraphics[trim=0cm 0cm 1cm 0.5cm, clip=true, width=0.4\textwidth]{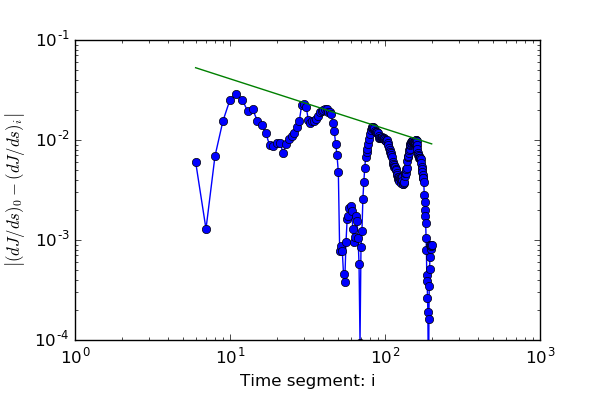}}	
		{\includegraphics[trim=0cm 0cm 1cm 0.5cm, clip=true, width=0.4\textwidth]{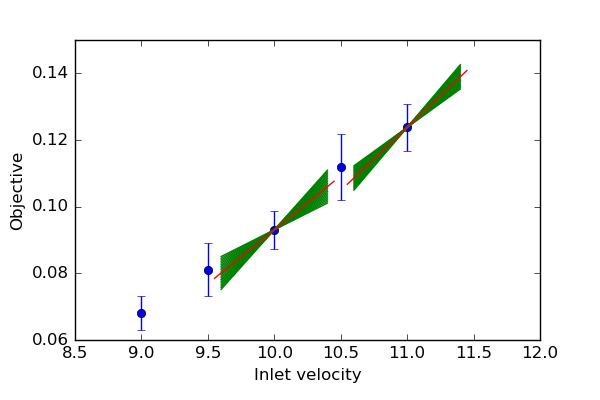}}
	\end{subfloat}\\
	\begin{subfloat}
		{\includegraphics[trim=0cm 0cm 1cm 0.5cm, clip=true, width=0.4\textwidth]{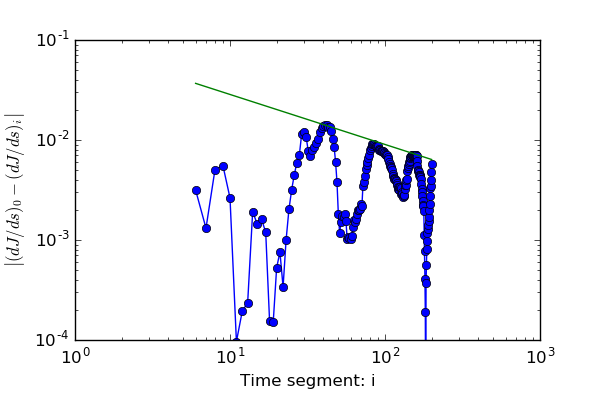}}
		{\includegraphics[trim=0cm 0cm 1cm 0.5cm, clip=true, width=0.4\textwidth]{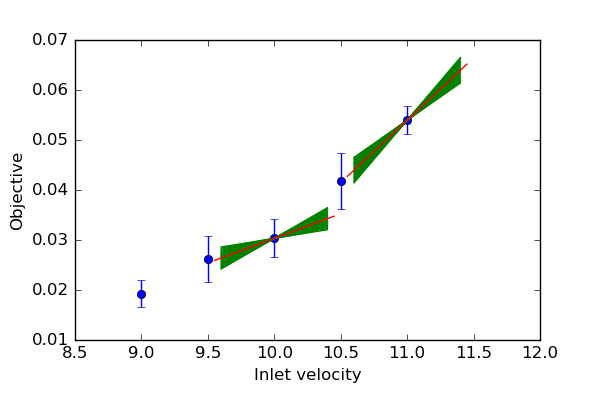}}
	\end{subfloat}\\
	\begin{subfloat}
		{\includegraphics[trim=0cm 0cm 1cm 0.5cm, clip=true, width=0.4\textwidth]{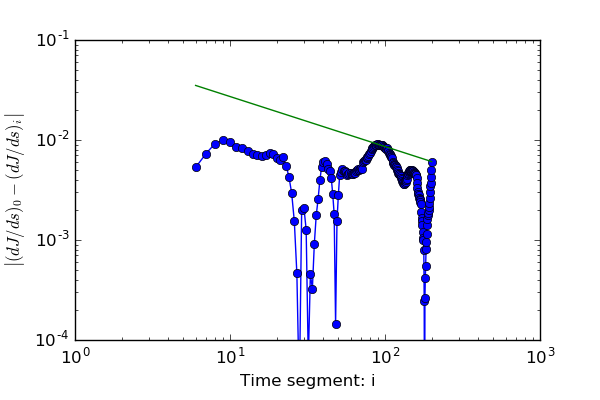}}
		{\includegraphics[trim=0cm 0cm 1cm 0.5cm, clip=true, width=0.4\textwidth]{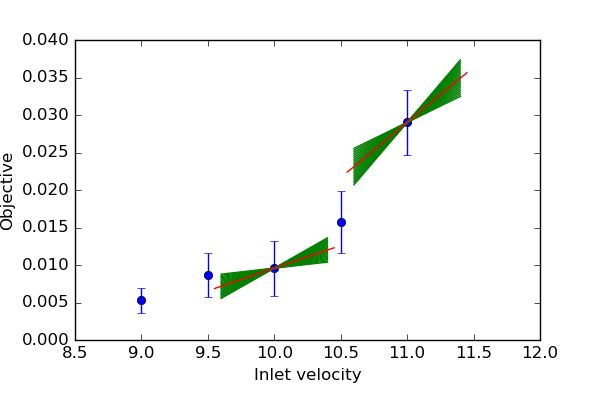}}
	\end{subfloat}\\
	\caption{Sensitivity computed by NILSS. From top to bottom, the objective function is the long-time average of $U_x/10$, $(U_x/10)^2$, $(U_x/10)^4$, and $(U_x/10)^8$. Left column: sensitivity computed by an increasing number of segments, the lines indicate confidence intervals for sensitivities. Right column: sensitivity plotted with objectives for adjacent parameters, the bars and wedges indicate confidence intervals of the objectives and sensitivities, respectively.}
	\label{f:pitzdaily dJds}
\end{figure}

Each objective $\avg{J}_\infty$ is approximated by $\avg{J}_{T'}$, which is the average of the instantaneous objectives $J(t)$ over $2\times 10 ^5$ time steps, or $T'=2$ seconds.
Since $J(t)$ exhibits aperiodic oscillations, $\avg{J}_{T'}$ has uncertainty.
To get the uncertainty, we divide the history of $J(t)$ into 5 equally long parts in time..
Denote the objectives averaged over each of the five parts by $J_1,...J_5$.
The corrected sample standard deviation between them are:
\begin{equation}
\sigma' = \sqrt{\frac{1}{4} \sum_{k=1}^{5} (J_k-\avg{J}_{T'})^2} .
\end{equation}
Here we assume that the standard deviation of $\avg{J}_{T'}$ is proportional to $T'^{-0.5}$. Thus, we use $\sigma = \sigma' / \sqrt{5}$ as the standard deviation of $\avg{J}_{T'}$.
We further assume $\pm2\sigma$ yields the 95\% confidence interval for $\avg{J}_{T'}$.
Objectives for different parameters in the range [9,11] are shown in the right column of figure \ref{f:pitzdaily dJds}, where the bars indicate the 95\% confidence interval.

Since we do not have tangent solvers, we use finite difference results to approximate all the tangent solutions used in NILSS:
this variant is called the finite difference NILSS (FD-NILSS), and it is discussed in detail in \cite{Ni_fdNILSS}.
\footnote{The python package `fds' implementing FD-NILSS is available at https://github.com/qiqi/fds. 
Comparing to the nilss package used in last subsection,
this code has better performance optimization and more sophisticate interfaces to existing solvers,
but might be more difficult for pedagogical purposes.
The particular files related to the application in this subsection are in fds/apps/openfoam4\_pitzdaily.}
For this FD-NILSS, we set each segment to have 250 time steps, or $\Delta T =  0.0025$ second.
To compute the sensitivity, we run NILSS over $K=200$ segments, or $T=0.5$ second.

\begin{figure}[htb]
	\centering
	\begin{subfloat}
		{\includegraphics[trim=0cm 0cm 0cm 0cm, clip=true, width=0.45\textwidth]{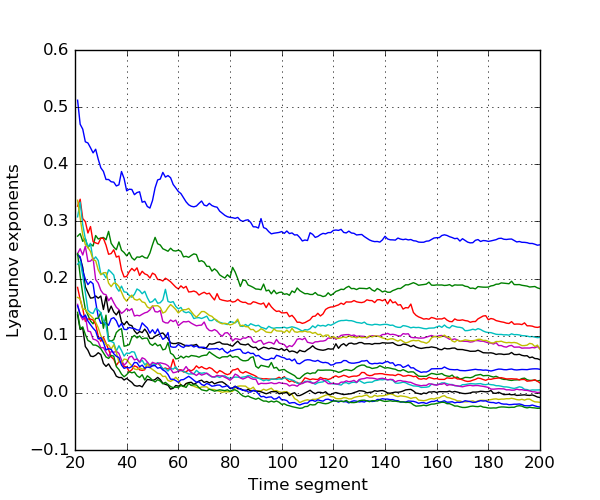}}
		{\includegraphics[trim=0cm 0cm 0cm 0cm, clip=true, width=0.45\textwidth]{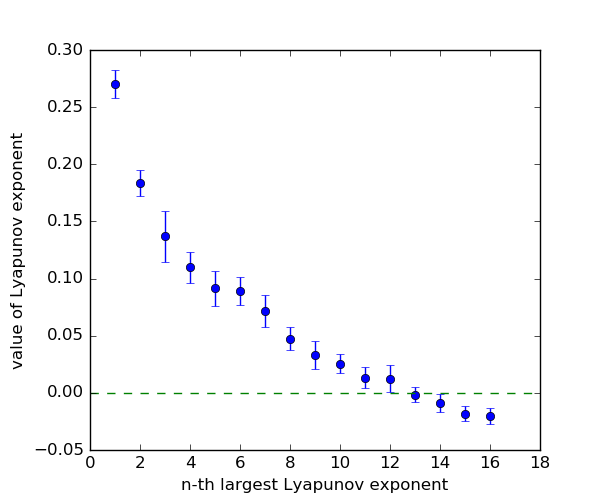}}
	\end{subfloat}\\
	\caption{Lyapunov exponents (LE). Left: the convergence history of 16 different LEs as the trajectory length increases, where the trajectory length is represented by the number of segments. Right: confidence interval of the largest 16 LEs. The unit of the y-axis is $\Delta T ^{-1} = 400$ second$^{-1}$. }
	\label{f:LE pitzdaily}
\end{figure}

To determine the number of homogeneous solutions, $M$, we compute LEs by the method described in section \ref{ss:determine parameters}.
For a particular LE, denoted by $\lambda$, its computed value changes with the length of the trajectory, or the number of segments, provided that the segment length $\Delta T$ is fixed.
We use $\lambda_i$ to denote the LE value computed using data from segments $1,2,...,i$.
To determine the uncertainty in the computed LE, we compute the smallest interval that converges at rate $i^{-0.5}$ and contains all $ \{\lambda_i\} $.
Specifically, we assume that $\{\lambda_i\}$ converges to some $\lambda_0$ as we increase $i$ and its confidence interval is proportional to $i^{-0.5}$.
To find $\lambda_0$, we first define $C(\lambda)$ as:
\begin{equation}
    C(\lambda) = \min\{C' \;|\; \left|\lambda - \lambda_i\right| \le C' i^{-0.5}, \text{for all } i\le K\},
\end{equation}
where $K$ is the number of segments.
We define $\lambda_0$ as such that the corresponding $C(\lambda_0)$ is smallest:
\begin{equation}
	\lambda_0 = \argmin_\lambda \{C(\lambda)\}.
\end{equation}
We regard $C K^{-0.5}$ as the confidence interval for $\lambda_0$.
The convergence history of the largest 16 LEs are shown in the left of figure \ref{f:LE pitzdaily}.
The $\lambda_0$ and confidence intervals for each LE are shown in the right of figure \ref{f:LE pitzdaily}.
The total number of positive LEs is smaller than 16.
So we set $M=16$.

By using the settings listed above, the cost of NILSS is mainly in integrating the primal solution over $200\times 250 \times 18 = 9 \times 10 ^5$ time steps.
Here $200$ is the number of segments, $250$ is the number of time steps in each segment, and $18$ is the number of primal solutions computed.
In finite difference NILSS, we need one $v^*$ and 16 $\{w_j\}$.
Each tangent solution is approximated by a finite difference between a perturbed solution and the same base solution: that is 18 primal solutions in total.

We want to give confidence intervals for the sensitivities computed by NILSS.
Similar to the case of LE, the value of $dJ/ds$ changes with $T$, or equivalently, the number of segments.
We use $(dJ/ds)_i$ to denote the sensitivity computed using data from segments $1,2,...,i$.
In this case, we assume that $\{(dJ/ds)_i\}$ converges to some $(dJ/ds)_0$ as we increase $i$, and its confidence interval is proportional to $i^{-0.5}$.
To find $(dJ/ds)_0$, we first define $C(dJ/ds)$ as
\begin{equation}
C\left(\dd Js\right) = \min\Bigg\{C' \;\Bigg|\; \left|\dd Js - \left(\dd Jss\right)_i\right| \le C' i^{-0.5}, \text{for all } i\le K\Bigg\}.
\end{equation}
We define $(dJ/ds)_0$ such that the corresponding $C((dJ/ds)_0)$ is the smallest:
\begin{equation}
\left(\dd Js\right)_0 = \argmin_{dJ/ds} \Bigg\{ C\left(\dd Js\right) \Bigg\}.
\end{equation}
We regard $C K^{-0.5}$ as the confidence interval for $(dJ/ds)_0$.
The left column in figure \ref{f:pitzdaily dJds} is a log-log plot of $\left|(dJ/ds)_0-(dJ/ds)_i\right|$ versus $i$ for $U_{x0}=10$, where the lines indicate $ C i^{-0.5}$.
Similarly, we find the confidence interval of the sensitivity at $U_{x0}=11$.
In the right column of figure \ref{f:pitzdaily dJds}, the wedges indicate the confidence intervals of the sensitivities.

As we can see in figure \ref{f:pitzdaily dJds}, in the last three rows, the sensitivities computed by NILSS correctly reflect the trend in long-time averaged objectives.
However, for the first row, the averaged objectives themselves have large uncertainties.
This is because a function oscillating near zero usually has large variance in comparison to its average.
In this scenario, since the primal simulation does not suggest a trend, we cannot tell if NILSS gives a meaningful derivative.

In our current example, the cost of NILSS is roughly the same as that of the conventional finite difference method.
For chaotic systems, with fixed $u_0$ and $T'$, the relation $\avg{J}_{T'} \sim s$ has many local fluctuations \cite{Wang_ODE_LSS}.
To smooth out these local fluctuations, we perform a linear regression over 5 parameters within the interval [9,11].
In the conventional finite difference method, the total cost comes from integrating the primal system for $5 \times 2 \times 10^5 = 1 \times 10 ^6$ steps.
This cost is similar to NILSS, which integrates for $9 \times 10 ^5$ steps.

However, here we may be making a comparison in favor of the conventional finite difference.
In figure \ref{f:pitzdaily dJds}, the range span of parameters is 2; it is too large for the last two objectives, since the relations between objectives and parameters are not linear.
In these cases, if we want to reduce the error in linearly approximating a nonlinear function, the parameter range should be smaller.
However, this requires the confidence intervals of the objectives to be reduced as well.
Otherwise, the uncertainties in the objectives are divided by a smaller parameter range; this would give rise to larger uncertainties in the sensitivities.
To obtain smaller confidence intervals for the objectives, we require longer trajectories, which means larger computational cost for the conventional finite difference method.

When there are multiple parameters, the cost of NILSS is even lower than the conventional finite difference method.
For a tangent NILSS, equation (\ref{eq:tangent equation}) has a right-hand side $\partial_s f$, which states that $v^*$ would change if we have a new parameter; however, $w_j$ does not depend on the parameter $s$, so they could be reused for the new parameter.
The marginal cost of adding a new parameter is only the cost to compute a new $v^*$.
In our finite difference NILSS for this problem, 18 trajectories were computed: one is a base trajectory, one has a perturbed parameter, 16 have perturbed initial conditions.
Only the trajectories with perturbed parameter should be recomputed for an additional parameter.
So the marginal cost of another parameter is only $1/18$ of the cost of the first parameter.
On the other hand, for the conventional finite difference method, 5 trajectories are computed: one is a base trajectory and 4 have perturbed parameters.
As a result, 4 trajectories should be recomputed for a new choice of parameter.
This suggests that the marginal cost of another parameter is $4/5$ of the cost of the first parameter, which is higher than that of finite difference NILSS.

The cost of NILSS is lower than LSS.
The number of states in our problem is $12225\times 3 = 36675$.
If we perform the conventional LSS over the same time span of $5 \times 10 ^4$ steps, the LSS method would require solving a linear equation system with $1.8\times 10^9$ variables.
This would be a very large cost in both computation time and computer storage.
In fact, comparing to the application of LSS on a airfoil \cite{Blonigan_lss_airfoil}, 
our application of NILSS in this paper uses a mesh with 6 times more cells,
the physical problem has 6 times more unstable CLVs,
we use a computer with only 1/60 many cores,
yet NILSS computes sensitivity 5 times faster.
Multiplying these factors together, we can roughly say that NILSS is thousands times faster than LSS on these open flow problems.

\section{Conclusions}

We develop the Non-Intrusive Least Squares Shadowing (NILSS) method
for computing the sensitivity of long-time averaged objectives of chaotic systems.
It has several advantages over LSS:
\begin{enumerate}
  \item NILSS explicitly exploits the CLV structure of tangent solutions, 
    and reduces the feasible set of LSS to a lower dimensional set affine to the unstable subspace.
    NILSS has low computational cost for problems with low dimensional unstable subspace, which is the case for many engineering applications.
  \item NILSS requires minor modifications to existing solvers. 
  \item NILSS consumes similar amount of computer memory as performing numerical simulation.
\end{enumerate}
NILSS has been demonstrated on the Lorenz 63 system and a CFD simulation for a flow over a backward-facing step.
For the latter case, NILSS is much faster than LSS, and has a similar computational cost as the numerical simulation.

\section*{Acknowledgment}
The authors acknowledge funding from AFOSR Awards FA9550-15-1-0072 under Dr. Fariba Fahroo and Dr. Jeanluc Cambrier,
AFOSR STTR Award FA9550-14-C-0024 under Dr. Phil Beran, and DOE Award DE-SC00011089.

\appendix

\section{Showing \texorpdfstring{$f(u)$}{f(u)} is a CLV with a zero LE} \label{ss:f is 0 CLV}

We assume the attractor $\Lambda$ is bounded with a positive lower bound for $f(u)$, i.e., there exists $C_1^0>0$, such that
\begin{equation}
\|f(u)\|\ge C^0_1, \quad \text{for all} \;u\in \Lambda.
\end{equation}
Since $f(u,s)$ is a continuous function, then $f(u)$ is continuous for fixed $s$. 
Together with the assumption that $\Lambda$ is bounded, we see that the $f(\Lambda)$ is also bounded, i.e., there exists $C^0_2>0$, such that
\begin{equation}
\|f(u)\|\le C^0_2, \quad \text{for all} \;u\in \Lambda.
\end{equation}

We check that for a fixed $s$, $f(u)$ is a homogeneous tangent solution that satisfies 
\begin{equation}
\dd {f(u)}{t} = \pp{f}{u} \dd{u}{t} = \partial_u f f,
\end{equation}
where the last equality is due to equation (\ref{eq:dynamical system}).

Next, we denote $C_1 = C^0_1/\|f(u(0))\|$, $C_2 = C^0_2/\|f(u(0))\|$, then $f(u)$ is a CLV whose LE is 0, since
\begin{equation}
C_1 e^{0 t }\|f(u(0))\|  \le \|f(u(t))\| \le C_2 e^{0 t }\|f(u(0))\|  ,
\end{equation}
which satisfies equation (\ref{eq:CLV}).

\section{Showing \texorpdfstring{$\{\|\zeta^\perp_j\|\}$} {zperp j} behave like exponentials} 
\label{ss: CLV perp are exponentials}

Here we show that the norm of the orthogonal projection of stable and unstable modes, $\{\|\zeta^\perp_j\|\}$, behave like exponentials.

We assume that all CLVs are uniformly bounded away from each other.
First, we define the angle $\alpha_{ij}(u)$ between two CLVs,
\begin{equation}
\alpha_{ij}(u) = \arccos \frac{\zeta_i(u)^T \zeta_j(u)}{\|\zeta_i(u)\| \|\zeta_j(u)\|}, \quad i\ne j.
\end{equation} 
The assumption means that there is an $\alpha _0>0$ such that:
\begin{equation}
\alpha_{ij}(u) >\alpha _0, \quad \text{for all} \; i\ne j, u\in \Lambda \;,
\end{equation}
where $\Lambda$ is the attractor.

Since $f(u)$ is also a CLV, the angles between $\{\zeta_j\}$ and $f(u)$ are all greater than $\alpha_0$ and the angles between $\{\zeta_j\}$ and $V^\perp$ are smaller than $\pi/2 -\alpha_0$.
Hence, by using the $C_1$ and $C_2$ provided by equation (\ref{eq:CLV}), we arrive at
\begin{equation}
\|\zeta_j^\perp(u(t))\| 
\ge \sin(\alpha_0) \|\zeta_j(u(t))\| 
\ge \sin(\alpha_0) e^{\lambda_j t}C_1\|\zeta_j(u(0))\| 
\ge C'_1 e^{\lambda_j t} \|\zeta_j^\perp(u(0))\| \;,
\end{equation}
where $C'_1 = \sin(\alpha_0) C_1$. 
On the other hand, we know that
\begin{equation}
\|\zeta_j^\perp(u(t))\| 
\le \|\zeta_j(u(t))\| 
\le C_2 e^{\lambda_j t}\|\zeta_j(u(0))\|  
\le C'_2 e^{\lambda_j t}\|\zeta^\perp_j(u(0))\| \;,
\end{equation}
where $C'_2 = \sin(\alpha_0) C_2$.
To summarize, there is $C_1', C_2' >0$, such that
\begin{equation}
C'_1 e^{\lambda_j t }\|\zeta^\perp_j(u(0))\|  
\le \|\zeta^\perp_j(u(t))\| 
\le C'_2 e^{\lambda_j t }\|\zeta^\perp_j(u(0))\|\;.
\end{equation}
Here all $\lambda_j\ne0$, since they correspond to either stable or unstable modes, but not the neutral mode.

\section{Derivation of \texorpdfstring{$d{\avg{J}_\infty}/{ds}$} {dJ/ds} } 
\label{ss:derive djds}

By applying an infinitesimal perturbation in $s$, the governing equation for $u$ is:
\begin{equation}
    \dd{(u+\delta u)}{t} = f(u+\delta u, s +\delta s)\;.
\end{equation}
After subtracting it by the unperturbed ODE, we get the governing equation for $\delta u$
\begin{equation}
	\dd{(\delta u)}{t} = \partial_u f \delta u + \partial _s f \delta s\;.
\end{equation}

\begin{figure}[htb]
    \centering
    \includegraphics[width=0.6\textwidth]{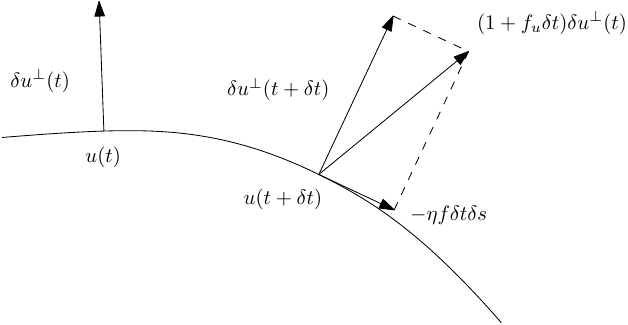}
    \caption{Perturbation of the trajectory due to a perturbation on the parameter.}
    \label{f:compare_trajec}
\end{figure}

As shown in fig \ref{f:compare_trajec}, we assume that at time $t$, the difference of the new trajectory from the original one is itself perpendicular to $f$, or $ \delta u(t) = \delta u^\perp (t)$.
After $\delta t$, this difference is no longer perpendicular to $f$, and thus it becomes 
\begin{equation}\label{eq:du by time step}
	\delta u (t+\delta t)= \delta u^\perp(t)  + (\partial_u f \delta u^\perp (t) + \partial _s f \delta s) \delta t \;.
\end{equation}

We denote the projection of $\delta u(t+\delta t)$ onto the direction of $f(u(t+\delta t))$ by $-\eta f \delta t \delta s$, or 
\begin{equation}
\begin{split}
-\eta f \delta t \delta s &= \frac
{f^T \left[\delta u^\perp(t+\delta t)\right]}
{f^T f} f   \;.
\end{split}
\end{equation}
On the other hand, the projection of $\delta u(t+\delta t)$ onto $V^\perp$ is denoted by $\delta u^ \perp (t+\delta t)$, as defined in equation (\ref{eq:vperp projection}).
Thus, in equation (\ref{eq:du by time step}), $ \delta u(t+\delta t)$ can be represented as the summation of two orthogonal projections:
\begin{equation}
\delta u^\perp(t)  + (\partial_u f \delta u^\perp (t) + \partial _s f \delta s) \delta t 
= \delta u^ \perp (t+\delta t) - \eta f \delta t \delta s\;.
\end{equation}
We recall our definition that $v= \delta u/ \delta s$, $v^\perp = \delta u ^\perp / \delta s$, we obtain:
\begin{equation} \label{eq:eta and v}
\dd{v^\perp}{t} = \partial_u f v^\perp + \partial_s f + \eta f\;.
\end{equation}
Here $v$ is the tangent solution of equation (\ref{eq:tangent equation}); $v^\perp$ is the orthogonal projection of $v$ according to equation (\ref{eq:vperp projection}).
Only $\eta$ is unknown, so we can also view equation (\ref{eq:eta and v}) as the definition of $\eta$.

Recall $\xi$ is the scalar such that $\xi f = v-v^\perp$, as defined in equation (\ref{eq:xi}).
We can show that:
\begin{equation} \label{eq:eta}
\eta = -\dd{\xi}{t}\;,
\end{equation}
To see this, first subtract equation (\ref{eq:eta and v}) from (\ref{eq:tangent equation}). This yields
\begin{equation}
\dd{(v - v^\perp)}{t} = \partial_u f (v-v^\perp) - \eta f\;.
\end{equation}
Using our definition of $\xi$, we arrive at
\begin{equation}
\dd{(\xi f)}{t} = \partial_u f (\xi f) - \eta f\;.
\end{equation}
By the rule for differentiating the product of two functions,
\begin{equation}
\dd{(\xi f)}{t} = \xi \dd{f}{t} + \dd{\xi}{t} f\;.
\end{equation}
Equation (\ref{eq:eta}) is obtained by recalling the chain rule for the differential:
\begin{equation}
\partial_u f (\xi f) = \xi \;\partial_u f (f)= \xi (\partial_u f \dd{u}{t}) = \xi \dd{f}{t}\;.
\end{equation}

To know the difference between the perturbed trajectory and the base trajectory, we need to define a correspondence between the states on the two trajectories.
That is, we should define which state on the base trajectory should be subtracted by which state on the perturbed trajectory.

Instead of comparing the two trajectories in the same time frame, we vary the length of infinitesimal time steps so that the corresponding states of the two trajectories remain perpendicular to $f$.
In time $\delta t$, the new trajectory moves a length of $f\delta t - \eta f \delta t \delta s$.
So the new speed is $(1-\eta \delta s)f$.
Hence the new trajectory needs time $\delta t/ (1-\eta \delta s) \approx \delta t (1+\eta \delta s) $ to cross length $f \delta t$, which is the length of the base trajectory.
If we compare the point on base trajectory at time $(t + \delta t)$ with the point on the perturbed trajectory at time $ t+ \delta t (1+\eta \delta s) $, their difference will remain perpendicular to $f$, which is $ \delta u^\perp (t + \delta t)$.

The $J_{new} \delta t_{new}$ on this small section of new trajectory is:
\begin{equation}
\begin{split}
&J_{new} \delta t_{new} \\
=&(J + \partial _u J \delta u^\perp) (1+\eta \delta s) \delta t \\
=&J\delta t + \partial _u J \delta u^\perp\delta t + J \eta \delta s \delta t\;.
\end{split}
\end{equation}

To compute the difference between the average $J$, we first write down its definition: 
\begin{equation}
\begin{split}
&\frac{1}{T_{new}} \int_{0}^{T_{new}} J_{new} dt  - \frac{1}{T} \int_{0}^{T} J dt\\
=& \frac{1}{\int_{0}^{T} (1+\eta \delta s) dt} \int_{0}^{T} \left(J + \partial _u J \delta u^\perp + J \eta \delta s \right) dt 	 - \frac{1}{T} \int_{0}^{T} J dt \\
=& \frac{\delta s}T \int_{0}^{T} \left[\partial_u J \, v^\perp+ \partial_sJ + \eta (J - \avg{J}) \right] \, dt\;,
\end{split}
\end{equation}
where we used the definition  $ \delta u^\perp (t) = v^\perp \delta s $.
If we divide by $\delta s$, we arrive at:
\begin{equation} \label{eq:dds finite T}
\dd{}{s} \left(\frac{1}{T} \int_{0}^{T} J dt\right) = \frac{1}T \int_{0}^{T} \left[\partial_u J \, v^\perp+ \partial_sJ + \eta (J - \avg{J}) \right] \, dt\;.
\end{equation}
Notice that here the ending time $T$ also depends on $s$.

First we use the shadowing direction $v^\infty$ as $v$ in equation (\ref{eq:dds finite T}). 
Since $v^{\infty\perp}(u)$ is uniformly bounded for all $u$ on the attractor, we can interchange the procedure of differentiating by $s$ and letting $T$ go to infinity:
\begin{equation} \label{eq:interchange limit}
\begin{split}
&\dd{}{s} \avg{J}_\infty = \dd{}{s} \left(\lim\limits_{T\rightarrow\infty}\frac{1}{T} \int_{0}^{T} J dt\right) \\
=&\lim\limits_{T\rightarrow \infty}\frac{1}T \int_{0}^{T} \left[\partial_u J \, v^{\perp\infty}+ \partial_sJ + \eta (J - \avg{J}) \right] \, dt	\;,
\end{split}
\end{equation}
where $ \eta $ is computed by substituting $v^\infty$ into equation (\ref{eq:eta and v}).
In fact, it is exactly the commutation between differentiation and $T$ going to infinity that requires $v^{\infty\perp}$ to be uniformly bounded.
The mathematical proof that justifies the interchange of two procedures can be found in \cite{wang2014convergence,Chater_convergence_LSS}.

For infinite $T$, only $v^{\infty\perp}$ can make \ref{eq:interchange limit} holds.
However, for finite $T$, we can use the NILSS solution $v$ to approximate $v^\infty$ and arrive at:
\begin{equation} 
\dd{\avg{J}_\infty}{s} \approx 
\frac 1T \int_{0}^{T} \left[\partial_u J \, v^{\perp}+ \partial_sJ + \eta (J - \avg{J}_T) \right] \, dt \;.	
\end{equation}
The proof of this approximation can be accomplished similarly to that in \cite{wang2014convergence,Chater_convergence_LSS}.

We can replace the requirement for computing $\eta$, by computing $\xi$ at the two ends of the trajectory.
To achieve this, we first apply equation (\ref{eq:eta}) and integrate by parts:
\begin{equation} 
\begin{split}\label{e:noname}
\dd{\avg{J}_\infty}{s} &\approx 
\frac 1T \int_{0}^{T} \left[\partial_u J \, v^{\perp}+ \partial_sJ -\dd{\xi}{t} (J - \avg{J}_T) \right] dt\\
&=\frac 1T \left[\int_{0}^{T} \left(\partial_u J \, v^{\perp}+ \partial_sJ\right) dt -\left(\xi J \right)\vert^T_0  + \xi \vert^T_0 \avg{J}_T + \int_{0}^{T}\xi \dd{J}{t} dt \right]
\end{split} .
\end{equation}
Next, we apply the fact that
\begin{equation}
\dd{J}{t} = \partial_u J \dd{u}{t} = \partial_u J \,f,
\end{equation}
and that $ v = v^\perp + \xi f $ into equation (\ref{e:noname}). Thus, we have:
\begin{equation} 
\begin{split}
\dd{\avg{J}_\infty}{s} &\approx \frac 1T \left[\int_{0}^{T} \left(\partial_u J \, v^{\perp}+ \partial_sJ + \xi \partial_u J f \right) dt 
-\left(\xi J \right)\vert^T_0  + \xi \vert^T_0 \avg{J}_T\right]\\
&=\frac 1T \left[
\int_{0}^{T} \left(\partial_u J \, v+ \partial_sJ \right)dt 
+\left. \xi\right\vert^T_0 \avg{J}_T
-\left. \left(\xi J \right)\right\vert^T_0\right]
\end{split} ,
\end{equation}
This is exactly equation (\ref{eq:derivative}).

\section{Derivation of \texorpdfstring{$ d{\avg{J}_\infty}/{ds} $} {dJds}  on multiple segments} 
\label{ss:derive djds multiple segments}

To derive equation (\ref{eq:dJds seg}) from equation (\ref{eq:derivative}), first we recover a continuous tangent solution $v$ from $\{v^{\perp}_i\}$ and $\{\xi_i\}$ on each segment:
\begin{equation}
v(t) = v^{\perp}(t) + \xi(t) f(t) \;, 
\end{equation}
where
\begin{equation}
\begin{cases}
v^\perp(t) = v_i^\perp(t), \\
\xi(t) = \xi_{i}(t) + \sum_{i'=0}^{i-1} \xi_{i'}(t_{i'+1}), 
\end{cases}
\quad t\in\left[t_i, t_{i+1}\right],
\end{equation}
where $ \{v_i^\perp(t)\} $ are given by equation (\ref{eq:v_i}), $\xi_{i}(t)$ are given by equation (\ref{eq:xi_i}).
The definition of $ \xi $ can be viewed as `accumulating' $\xi_i$ from previous segments.
Applying this definition, we have:
\begin{equation}\label{e:xi0T}
\xi(0)=0, \quad \xi(T)=\sum_{i=0}^{K-1}\xi_i(t_{i+1}) .
\end{equation}

The continuity of $v$ follows from the continuity of $v^{\perp}$ and $\xi$.
$v^\perp(t)$ is continuous because of the continuity condition in equation (\ref{eq:cts requirement}).
$\xi(t)$ is continuous because $\xi_{i}(t_i)=0$, as shown in equation (\ref{eq:xi end value}).

To see that $v$ is a tangent solution of equation (\ref{eq:tangent equation}), we first notice that on segment $i$, v(t) is characterized by 
\begin{equation}
v(t) = v_i(t) + \xi^*_i f(t) ,\quad t\in\left[t_i, t_{i+1}\right],
\end{equation}
where $\xi^*_i = \sum_{i'=0}^{i-1} \xi_{i'}(t_{i'+1}) $.
Taking the time derivative of $v$, we have:
\begin{equation}
\begin{split}
\dd{v}{t} &= \dd{v_i}{t} + \xi^*_i \dd{f}{t} = \partial_u f v_i + \partial_s f + \xi^*_i \partial_u f f\\
&=  \partial_u f (v_i+\xi^*_i f) +\partial_s f = \partial_u f v +\partial_s f . 
\end{split}
\end{equation}

To conclude, $v$ is a continuous tangent solution over the entire trajectory and the $L^2$ norm of $v^\perp$ is minimized, i.e., $v$ is the solution of NILSS problem on a single time segment.
Hence we can substitute $ v $ into equation (\ref{eq:derivative}), which means that, together with equation (\ref{e:xi0T}), we obtain:
\begin{equation}
\begin{split}
\dd{\avg{J}_\infty}{s} &\approx 
\frac 1T \left[
\int_{0}^{T} \left(\partial_u J \, v+ \partial_sJ \right)dt 
+\left. \xi\right\vert^T_0 \avg{J}_T
-\left. \left(\xi J \right)\right\vert^T_0\right]\\
&=\frac 1T \left[\sum_{i=0}^{K-1} 
\int_{t_i}^{t_{i+1}} 
\left(\partial_u J \, v_i+ \partial_sJ + \xi^*_i \partial_u J  \, f\right)dt \right]
+ \frac{1}{T}\left[\xi(T) (\avg{J}_T - J(T))\right]\\
&=\frac 1T \left[\sum_{i=0}^{K-1} 
\int_{t_i}^{t_{i+1}} 
\left(\partial_u J \, v_i+ \partial_sJ + \xi^*_i \dd{J}{t} \right)dt \right]
+ \frac{1}{T}\left[\xi(T) (\avg{J}_T - J(T))\right]\\
&=\frac 1T\sum_{i=0}^{K-1} \left[
\int_{t_i}^{t_{i+1}} 
\left(\partial_u J \, v_i+ \partial_sJ \right)dt 
+ \xi_i(t_{i+1}) (\avg{J}_T - J(t_{i+1})) \right].
\end{split}
\end{equation}
This yields equation (\ref{eq:dJds seg}).

\renewcommand*{\bibfont}{\small}
\bibliographystyle{model1-num-names}
\bibliography{MyCollection}

\end{document}